\documentclass[superscriptaddress,aps,prd,reprint]{revtex4-2}

\usepackage{lipsum}
\usepackage{blindtext}
\usepackage{graphicx}

\usepackage{dcolumn}
\usepackage{bm}

\usepackage{lipsum}  

\usepackage[utf8]{inputenc}
\usepackage[T1]{fontenc}

\usepackage{hyperref}
\usepackage{slashed}
\usepackage{amsfonts}
\usepackage{mathrsfs}
\usepackage{color}
\usepackage[makeroom]{cancel}
\usepackage{graphicx}
\usepackage{amssymb}
\usepackage{amsmath}
\usepackage{hyperref}
\usepackage{cleveref}
\usepackage{enumitem}

\renewcommand{\d}{\textrm{d}}

\def\p{\partial}

\def\bi{\begin{itemize}}
\def\ei{\end{itemize}}
\def\be{\begin{equation}}
\def\ee{\end{equation}}
\newcommand{\bea}{\begin{eqnarray}}
\newcommand{\eea}{\end{eqnarray}}

\begin{document}


\title{Vacuum Amplification of Chiral Gravitational Waves\\ and the Stochastic Gravitational Wave Background}

\author{Stephon Alexander}
\email{stephon\_alexander@brown.edu}

\author{Heliudson Bernardo}
\email{heliudson\_bernardo@brown.edu}

\author{Yiya Selina Li}
\email{yiya\_li@alumni.brown.edu}

\author{Cooper Niu}
\email{cooper\_niu@brown.edu}

\affiliation{Department of Physics, Brown University, Providence, RI 02912, USA}


\begin{abstract}
\noindent
We investigate cosmological vacuum amplification of gravitational waves in dynamical Chern-Simons gravity. We develop a comprehensive framework to compute graviton production induced by the parity violating Pontryagin coupling and study its imprint on the stochastic gravitational wave background energy power spectrum. We explore gravitational vacuum amplification in four concrete scenarios for the evolution of the Chern-Simons pseudoscalar. We show that a parity-violating contribution dominates over an initially flat spectrum when the velocity of the pseudoscalar quickly interpolates between two asymptotically constant values or when it is nonvanishing and constant through a finite period of time. This is also the case when we parametrize the pseudoscalar evolution by a perfect fluid with radiation- and dust-like equations of state for large enough values of its energy density. The resulting spectra are compared with the sensitivity curves of current and future gravitational wave observational searches.
\end{abstract}

\maketitle

\section{Introduction}
\noindent
The discovery of parity violation in the weak interaction has been pivotal in shaping the Standard Model (SM) of particle physics \cite{PhysRev.104.254}. As the parity violation in weak sector might originate from some high-energy or UV theory, it has inspired searches for new physics beyond the Standard Model that exhibits similar or related violations. Recent evidence of parity violation in the four-point galaxy correlation function \cite{Philcox:2022hkh, Hou:2022wfj} and Planck's EB angular power spectrum \cite{Minami:2020odp, Eskilt:2022cff} suggest a gravitational sector that violates parity. Although General Relativity (GR) is a parity-even theory, parity violation emerges as a generic prediction for many modified theories of gravity. Notable examples include Chern-Simons theory \cite{Alexander:2009tp, Jackiw:2003pm}, ghost-free scalar-tensor theory \cite{PhysRevD.97.044034, PhysRevD.101.024002}, teleparallel gravity \cite{Conroy:2019ibo}, and Horava-Lifshitz theory \cite{PhysRevD.79.084008, PhysRevD.88.063508}. 

Chern-Simons (CS) modified gravity is among the most studied parity-violating theories of gravity. With strong motivation from particle physics \cite{Alvarez-Gaume:1983ihn} and string theory \cite{Alexander_2006}, Chern-Simons theory extends GR with a coupling term between a gravitational Chern-Simons term and a (pseudo) scalar field. The CS term induces an asymmetry between the left- and right-handed gravitational wave (GW) polarizations. During gravitational wave propagation, parity violation amplifies one chirality and suppresses the other one (see \textit{e.g.} \cite{Jenks:2023pmk}). Birefringence can also lead to phase shift and thus modify the dispersion relation. As they might leave imprints on a variety of cosmological and astrophysical observables (\textit{e.g.} \cite{Manton:2024hyc,Yunes:2010yf, Alexander:2007kv, Gluscevic:2010vv, Lue:1998mq}), birefringent effects are a powerful probe for parity violation, especially during the ongoing era of GW astrophysics. 

The \textit{stochastic gravitational wave background} (SGWB) arises from a multitude of astrophysical and cosmological sources such as supermassive black hole mergers \cite{1995ApJ...446..543R, NANOGrav:2023hfp, EPTA:2023xxk}, cosmic strings \cite{Accetta:1988bg, Ellis:2020ena, Cui:2017ufi}, early universe phase transitions \cite{Gouttenoire:2023bqy, Addazi:2023jvg}, and inflation \cite{Starobinsky:1979ty, Vagnozzi:2023lwo}, creating a rich tapestry of signals that are sensitive to modifications of GR. Recent pulsar timing array (PTA) measurements have suggested the presence of SGWB at nanohertz frequencies \cite{NANOGrav:2023gor, EPTA:2023fyk, Reardon:2023gzh}. PTAs utilize stable and fast rotating millisecond pulsars as precise astrophysical clocks to monitor small fluctuations in the spacetime metric. By computing the angular correlations across the sky, PTAs present exceptional sensitivity to detect faint and long-wavelength gravitational waves and constrain their amplitude and energy density.

Throughout cosmic history, the Chern-Simons pseudoscalar $\varphi$ might acquire different values either through its potential or couplings with other matter fields. A cosmological evolution of $\varphi$ is complementary to its local evolution, which is dominated by local curvature, such as in black-hole and stellar solutions (see \textit{e.g.} \cite{Ali-Haimoud:2011wpu,Yagi:2012ya} for explicit examples). For example, one can imagine a rolling solution for $\varphi$ right after the Hubble scale gets slightly smaller than the curvature scale of its potential. This situation is natural if $\varphi$ is thought of as an axion- or pion-like field. Accordingly, the field value and rolling speed vary and lead to nontrivial gravitational effects. As we shall see, depending on the field's velocity, the $\varphi$ evolution induces a non-adiabatic shift in the vacuum states of certain metric fluctuation modes, leading to the production of gravitons. Hence, the cosmological evolution of the pseudoscalar might leave an imprint on the SGWB. 

In the context of a Friedmann-Lema\^itre-Robertson-Walker (FLRW) background in GR, similar effects have been extensively studied \cite{Parker:1969au, Parker:1971pt, Grishchuk:1974ny, Ford:1986sy, Moorhouse:1994nc}. In inflationary cosmology, particle production is the mechanism that gives origin to the large-scale structure of the universe. A sudden scale-factor transition, for example, from a quasi-de-Sitter spacetime to a radiation-dominated epoch, can break the adiabatic evolution of the vacuum state for modes whose frequency is smaller than the inverse of the time scale associated with the transition \cite{Maggiore:1999vm}.

In this paper, we generalize the cosmological vacuum amplification mechanism to include scalar-induced gravitational wave production. For concreteness, we focus on Chern-Simons modified gravity, but we emphasize that such vacuum amplification is applicable to other scalar-tensor theories. The essential idea is that the cosmological evolution of $\varphi$ modifies the graviton mode function and thus the vacuum for the tensor modes of the metric, such that an initial vacuum state gets excited after $\varphi$ changes. As we will discuss, the effect is present even in flat space, but it also modifies the usual production of gravitational waves in cosmology. We aim to provide a comprehensive framework for studying vacuum amplifications of parity-violating gravitational waves and determine the parity violation imprinted on the cosmological SGWB. One application of our formalism is the possibility of using SGWB observations to investigate the cosmological evolution of $\varphi$ in a model-independent way (since we do not need to assume any couplings to the SM fields).

The paper is organized as follows. In Sec.~\ref{sec:vacuum_amp}, we review the vacuum amplification due to cosmic evolution in the context of GR. We then introduce the formalism of Chern-Simons-induced vacuum amplification in Sec.~\ref{sec:vacuum_amp_CS} and investigate the produced SGWB signal in four scenarios in Sec.~\ref{sec:application}. Lastly, we conclude in Sec.~\ref{sec:conclusion}. Our convention for the metric signature is $(-,+,+,+)$, and we work with natural units $c = \hbar = 1$ when not otherwise stated. Throughout the paper, we use boldface ($\mathbf{k}$) to denote spatial three-vector, circumflex accents ($\hat{k}$) for unit vectors, and $(k \equiv |\mathbf{k}|)$ for the magnitudes of the three-vectors. The overhead dot represents cosmic time derivatives $(\dot ~) \equiv (\partial_t)$, and prime represents conformal time derivative $(')\equiv(\partial_\eta)$.

\section{Vacuum Amplification in \\ General Relativity} \label{sec:vacuum_amp}

\subsection{Bogoliubov Transformation}
We proceed in the Friedmann-Lemaître-Robertson-Walker (FLRW) background to obtain cosmological gravitational wave solutions. The metric of interest is given by 
\begin{align}
   \d s^2 = a^2(\eta) \left[-\d\eta^2 + (\delta_{ij}+h_{ij}(\eta,{\bf x}))\d x^i \d x^j\right],
\end{align}
where we choose $h_{ij}$ to be in the transverse and traceless~(TT) gauge, \textit{i.e.,} $h_{ii} = \partial^j h_{ij} = 0$, and $a(\eta)$ the scale factor as a function of the conformal time $\eta$. The metric perturbation can be expanded as
\begin{align}\label{GW_expansion}
    \begin{aligned}
        h_{ij}(\eta,\mathbf{x})  &= \sqrt{16\pi G}\sum_{\lambda} \int \frac{\d^3 k}{(2 \pi)^3}\frac{\mathcal{Q}_\lambda(\eta, \mathbf{k})}{\sqrt{2k}} e_{ij}^\lambda(\hat{k}) e^{i\mathbf{k}\cdot \mathbf{x}}, \\
        \mathcal{Q}_\lambda(\eta, \mathbf{k})&= \hat{a}_\lambda(\mathbf{k}) \frac{\xi_{\mathbf{k}}(\eta)}{a(\eta)} +\hat{a}_\lambda^{\dagger}(-\mathbf{k}) \frac{\xi_{-\mathbf{k}}^*(\eta)}{a(\eta)},
    \end{aligned}
\end{align}
where $\lambda = \{+, \times\}$ labels the GW polarizations, and the two polarization tensors are defined in terms of the principal axes $\hat{m}$ and $\hat{n}$ that are perpendicular to the GW propagation direction $\hat{k}$, 
\begin{subequations}
    \begin{align}
    e_{ab}^+(\hat{k}) & = \hat{m}_a\hat{m}_b-\hat{n}_a\hat{n}_b ,\\
    e_{ab}^\times(\hat{k}) &= \hat{m}_a\hat{m}_b + \hat{n}_b\hat{n}_a.
\end{align}
\end{subequations}
From linearized Einstein's equation, with the choice of normalization \eqref{GW_expansion}, the creation and annihilation operators $\hat{a}^\dagger_\lambda(\mathbf{k})$ and $\hat{a}_\lambda(\mathbf{k})$ satisfy the canonical commutation relation 
\begin{equation}
    \left[ \hat{a}_\lambda(\mathbf{k}),\hat{a}^\dagger_{\tilde{\lambda}}(\mathbf{\tilde{k}})\right] = (2\pi)^3\delta_{\lambda \tilde{\lambda}}\delta^3(\mathbf{k}-\mathbf{\tilde{k}}), 
\end{equation}
with other commutators vanishing, whereas the mode function $\xi_{\mathbf{k}}(\eta)= \xi_{k}(\eta) $ satisfies
\begin{align}\label{eq:mode_func}
    \xi_{k}'' + \left(k^2 - \frac{a''}{a}\right)\xi_{k} = 0.
\end{align}
A key distinction of quantum field theory in curved space, as opposed to flat space, is that the choice of mode function is not unique. In Minkowski spacetime, a timelike Killing vector (which can be thought of as $\p_t$ in inertial coordinates) is associated with time translation invariance, which guarantees a unique choice of mode function with positive frequency. However, such a Killing vector no longer exists in a general curved spacetime. Diffeomorphism invariance excludes a preferred choice of time and, consequently, the preferred choice of mode functions \cite{Birrell:1982ix}.

Consider two cosmological epochs with distinct scale factors. The effective time-dependent frequency in \eqref{eq:mode_func} will be different in the two eras, and therefore the vacuum state will not be preserved by time evolution, giving origin to particle production. In the Heisenberg picture, the annihilation and creation operators for a given era $\hat{a}_\lambda(\mathbf{k})$ and $\hat{a}^\dagger_\lambda(\mathbf{k})$ can be rewritten as a canonical transformation of the annihilation and creation operators $\hat{A}_\lambda(\mathbf{k})$ and $\hat{A}^\dagger_\lambda(\mathbf{k})$ from the previous era, while preserving the commutation relations. The relation is given by a Bogoliubov transformation of the form \cite{Maggiore:2018sht}
\begin{align}\label{Bugoliubovdecomp}
    \hat{a}_\lambda (\mathbf{k}) = \alpha_\lambda(k) \hat{A}_\lambda (\mathbf{k}) + \beta^*_\lambda(k) \hat{A}^\dagger_\lambda (-\mathbf{k}),
\end{align}
where there is no sum on the right-hand side and $\alpha_\lambda(k)$ and $\beta_\lambda(k)$ are called Bogoliubov coefficients, which can be fixed by demanding the continuity of the mode functions across the transition between the two epochs. Due to the symmetries of the background, the Bogoliubov transformation can only depend on the modulus $k$ and there is no mix of different polarizations because they decouple in GR. The latter is also true in dCS provided we work with circular polarizations \cite{Isi:2018miq}. 
\begin{figure}[h]
    \centering
    \includegraphics[width=0.9\linewidth]{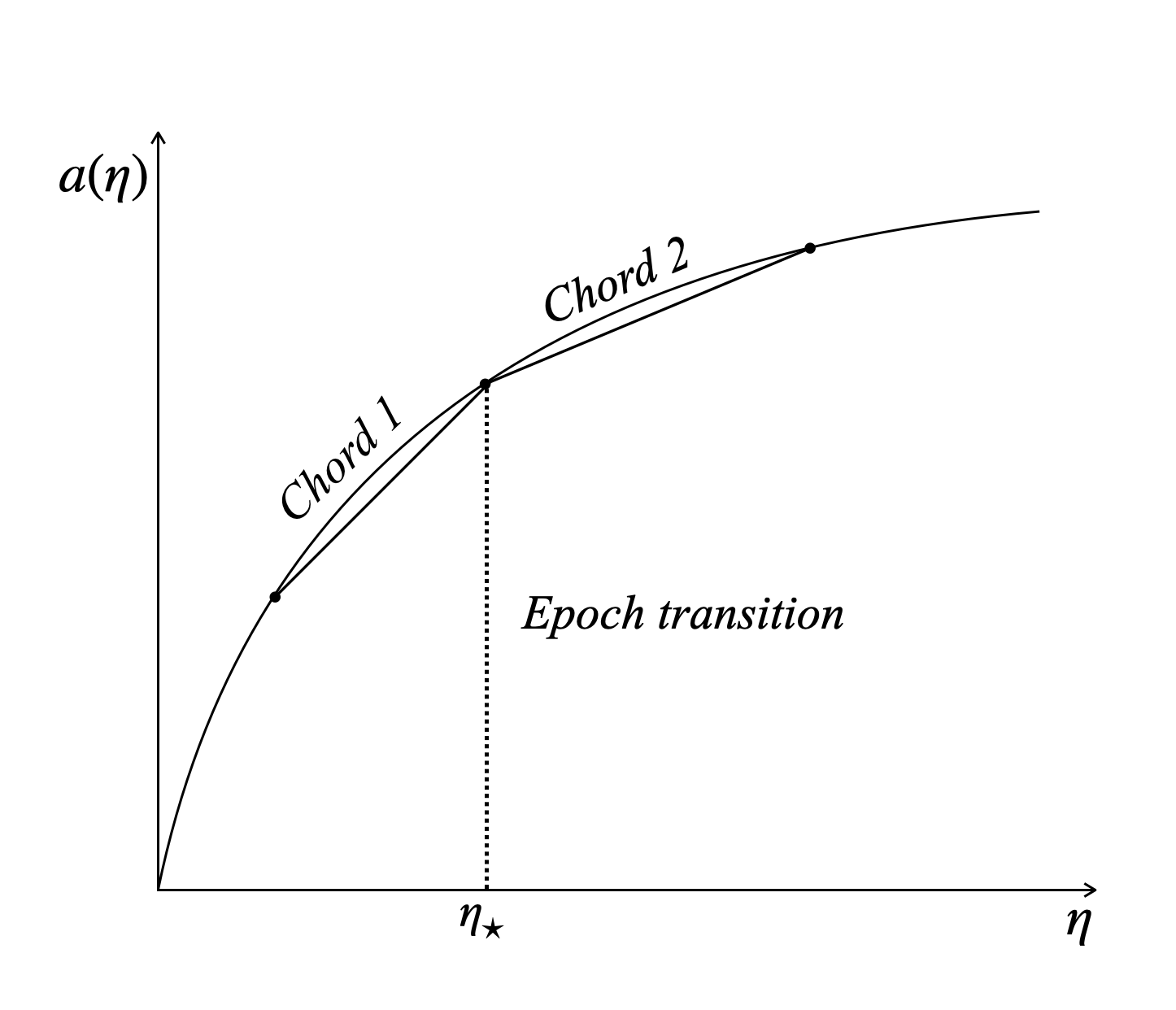}
    \caption{The graph of scale factor $a(\eta)$ can be approximated as segments of straight lines. For dCS gravity, the scale factor is replaced by $z(\eta)$ (see section \ref{sec:bog_coeff_dCS}).}
    \label{fig1:epoch_transition}
\end{figure}

Although a single Bogoliubov transformation is enough to relate the creation and annihilation operators between a sudden transition of the scale factor \cite{Abbott:1985cu}, we wish to have a formalism that can continuously track particle production as the scale factor evolves. This is achieved within the framework of \cite{Moorhouse:1994nc, Mendes:1994ai}, which we briefly review in the rest of this subsection.

In  \cite{Moorhouse:1994nc, Mendes:1994ai} the scale factor $a(\eta)$ is approximated by segments of straight lines, also called chords, where each segment is described by 
\begin{align}\label{eq:segment}
    a_n(\eta) \simeq B_n(\eta - b_n),
\end{align}
where $B_n$ and $b_n$ are the slope and intercept in the $n$-th segment. Consider two neighboring segments $1$ and $2$ transiting at $\eta = \eta_\star$ (see Fig. \ref{fig1:epoch_transition}). We can find the Bogoliubov coefficients by requiring the metric perturbation $h_{ij}$ and its first derivative to be continuous at $\eta = \eta_\star$:
\begin{subequations}
    \begin{align}
    &\alpha_1 \frac{\xi_1}{a_1} + \beta_1 \frac{\xi_1^*}{a_1}= \alpha_2 \frac{\xi_2}{a_2} + \beta_2 \frac{\xi_2^*}{a_2}, \\
    &\begin{aligned}
        & \alpha_1\left(\frac{\xi_1'}{a_1} - \frac{a_1'}{a_1^2}\xi_1\right) + \beta_1\left(\frac{{\xi_1^*}'}{a_1} - \frac{a_1'}{a_1^2}\xi_1^*\right) \\
        & = \alpha_2\left(\frac{\xi_2'}{a_2} - \frac{a_2'}{a_2^2}\xi_2\right) + \beta_2\left(\frac{{\xi_2^*}'}{a_2} - \frac{a_2'}{a_2^2}\xi_2^*\right).
    \end{aligned}
\end{align}
\end{subequations}
Here, we drop the polarization subscript $\lambda$ to avoid cluttering. Using the segment approximation, we have
\begin{subequations}
    \begin{align}
    \alpha_1 \xi_1 + \beta_1 \xi_1^* &= \alpha_2\xi_2 + \beta_2 \xi_2^*, \\
    \alpha_1 \xi_1' + \beta_1 {\xi_1^*}' + \rho\left(\alpha_1 \xi_1 + \beta_1 \xi_1^*\right) &=\alpha_2 \xi_2' + \beta_2 {\xi_2^*}'
\end{align}
\end{subequations}
with $\rho \equiv (\eta_\star - b_2)^{-1} - (\eta_\star - b_1)^{-1}$
Due to the straight line segment approximation in eq.~(\ref{eq:segment}), $a''/a$ in eq.~(\ref{eq:mode_func}) vanishes. Thus, the solution for the mode function is $\xi(\eta) \sim e^{-ik(\eta - \tilde{\eta})}$ with $\tilde{\eta}$ the same constant for all segments. The relation between the Bogoliubov coefficients in chord 2 and chord 1 are then solved by
\begin{subequations}
    \begin{align}
        & \alpha_2=\alpha_1+\frac{i}{2 k}\left(\alpha_1+\beta_1 e^{2 i k\left(\eta_\star-\tilde{\eta}\right)}\right) \rho, \\
        & \beta_2=\beta_1-\frac{i}{2 k}\left(\beta_1+\alpha_1 e^{-2 i k\left(\eta_\star-\tilde{\eta}\right)}\right) \rho.
    \end{align}
\end{subequations}
In the limit of infinitely short segments, the Bogoliubov coefficients become continuous and satisfy \cite{Moorhouse:1994nc}
\begin{subequations}
    \begin{align}
        & \alpha_\lambda^{\prime}(\eta)=\frac{i}{2 k}\left[\alpha_\lambda(\eta)+\beta_\lambda(\eta) e^{2 i k\left(\eta-\tilde{\eta}\right)}\right] \frac{a^{\prime \prime}(\eta)}{a(\eta)} \label{eq:alpha_prime}, \\
        & \beta_\lambda^{\prime}(\eta)=-\frac{i}{2 k}\left[\beta_\lambda(\eta)+\alpha_\lambda(\eta) e^{-2 i k\left(\eta-\tilde{\eta}\right)}\right] \frac{a^{\prime \prime}(\eta)}{a(\eta)}  \label{eq:beta_prime}.
    \end{align}
\end{subequations}
Given a scale factor $a(\eta)$ this system can be solved and the final particle number obtained, as we shall see in the next subsection. For more details on the continuous Bogoliubov coefficients formalism, see \cite{Moorhouse:1994nc, Mendes:1994ai}.

We can also parameterize the Bogoliubov coefficients in terms of two new variables $X$ and $Y$ as
\begin{subequations}\label{eq:ab_to_XY}
\begin{align}
\alpha_\lambda(\eta)&=\frac{1}{2}\left(X_\lambda+Y_\lambda\right) e^{i k \eta} \label{eq:alpha_XY},\\
\beta_\lambda(\eta)&=\frac{1}{2}\left(X_\lambda-Y_\lambda\right) e^{-i k \eta} \label{eq:beta_XY}.
\end{align}
\end{subequations}  
Eqs.~(\ref{eq:alpha_prime}) and (\ref{eq:beta_prime}) are then equivalent to
\begin{subequations}
    \begin{align}
        & X_\lambda^{\prime \prime}+\left(k^2-\frac{a''}{a}\right) X_\lambda = 0, \label{eq:x_a}\\
        & Y_\lambda = \frac{i}{k} X_\lambda' \label{eq:y_a}.
    \end{align}
\end{subequations}  
Notice that eq.~(\ref{eq:x_a}) is nothing but the Mukhanov–Sasaki equation and thus can be simply solved in different cosmic epochs. Combining with eq.~(\ref{eq:y_a}), we can retrieve the continuous Bogoliubov coefficients $\alpha_\lambda$ and $\beta_\lambda$ reversely.

\subsection{Energy Density and the primordial power spectrum}
In GR, the energy density of the stochastic gravitational wave background is given by the Isaacson formula \cite{Isaacson:1968hbi, Isaacson:1968zza}, which in the TT gauge reads
\begin{equation}\label{eq:stress_tensor_gr}
    T_{\mu\nu} = \frac{1}{32\pi G}\langle \langle \nabla_\mu h_{\alpha\beta}\nabla_\nu h^{\alpha\beta} \rangle\rangle,
\end{equation}
where the brackets stand for averages over the frequency of the GW oscillations. Hence, inside the Hubble radius and in conformal time coordinate, we have
\begin{equation}
    \rho_{\rm GW} = T_{00} = \frac{1}{32\pi G a^{2}(\eta)}\langle \langle h_{ij}' h_{ij}'\rangle \rangle.
\end{equation}
Using the mode decomposition \eqref{GW_expansion} gives
\begin{align}
    \rho_{\rm GW} &= \frac{1}{2a^{2}(\eta)}\sum_{\lambda, \tilde{\lambda}}
    \int \frac{\d^3{\bf k}}{(2\pi)^3}\frac{\d^3{\bf\tilde{k}}}{(2\pi)^3}\frac{e^\lambda_{ij}(\hat{k})e^{\tilde{\lambda}ij}(\hat{\tilde{k}})}{2\sqrt{k \tilde{k}}} \nonumber\\
    &\hspace{2.2cm}\times \langle \mathcal{Q}_\lambda^{\dagger\prime} (\eta,-\mathbf{k}) \mathcal{Q}_{\tilde{\lambda}}' (\eta,\mathbf{\tilde{k}})\rangle e^{i(\mathbf{k}+\mathbf{\tilde{k}})\cdot \mathbf{x}},
\end{align}
where we assumed ergodicity and late-time classicalization of the quantum fluctuations \cite{Isi:2018miq}.

To compute the power spectrum and the energy density of the GW background, we need $\langle \mathcal{Q}_\lambda'(\eta,\mathbf{k})  \mathcal{Q}_{\tilde{\lambda}}'(\eta,\mathbf{\tilde{k}})\rangle$, where the state might have a non-trivial occupation number at some initial time. This correlation function depends on the functional form of the mode function. However, we are interested in the GW background on sub-Hubble scales today because only sub-Hubble modes are observable. For $k \eta \gg 1$, we have
\begin{align}\label{eq:Q_correlation}
    \langle \mathcal{Q}_\lambda^{\dagger\prime} (\eta,-\mathbf{k})\mathcal{Q}_{\tilde{\lambda}}' (\eta,\mathbf{\tilde{k}})\rangle \simeq \frac{k\tilde{k}}{a^2(\eta)}\delta_{\lambda \tilde{\lambda}}(2\pi)^3 \delta^{(3)}(\mathbf{k}+\mathbf{\tilde{k}})\nonumber\\
    \times\left(|\alpha_\lambda|^2 + |\beta_\lambda|^2 - 2 \text{Re}\left[\alpha_\lambda \beta_\lambda^* e^{-2ik \eta}\right]\right)(2N_\lambda(k)+1),
\end{align}
where $N_{\lambda}(\mathbf{k})$ is the number operator associated to the initial creation operator $\hat{A}^\dagger_\lambda(\mathbf{k})$.
Here we assume the initial background to be isotropic, which means that the number density operator satisfies $N_\lambda(\mathbf{k})= N_\lambda(k)$. Note that even if the initial state is the vacuum state with $N_\lambda(\mathbf{k})=0$, there might be a non-trivial contribution to the two-point function at late times. 

Moreover, since only spacetime averages of the derivatives of $h_{ij}$ enter the Isaacson formula, the oscillatory term in eq.~(\ref{eq:Q_correlation}) will not contribute to the energy density. Using $|\alpha_\lambda|^2 -|\beta_\lambda|^2 = 1$, we obtain the following result
\begin{equation}\label{rhoGW}
    \rho_{\rm GW} = \frac{1}{4\pi^2 a^{4}(\eta)} \sum_\lambda \int \d k k^3\left(1+2|\beta_\lambda|^2\right)\left(2N_\lambda(k)+1\right).
\end{equation}
This should be compared with the energy density calculated from a phase space occupation number $n_\lambda(k)$
\begin{equation}
    \rho_{\rm GW} = \frac{1}{4\pi^2a^4}\sum_\lambda \int \d k k^3 (2n_{\lambda}(k)+1),
\end{equation}
where the last term in the bracket is the usual vacuum contribution to the energy density of any bosonic (and canonically normalized) quantum field.
Thus, we find
\begin{equation}
    n_\lambda = N_\lambda + |\beta_\lambda|^2(2N_\lambda +1),
\end{equation}
which could also be computed from the Bogoliubov transformation between the initial and final creation and annihilation operators \cite{Maggiore:1999vm}. 

The energy power spectrum produced by the vacuum amplification mechanism is then
\begin{equation}
    \rho_{\rm GW}(k) = \frac{\d\rho_{\rm GW}}{\d \ln k} = \frac{1}{2\pi^2 a^4(\eta)}\sum_\lambda k^4 n_\lambda(k),
\end{equation}
where only sub-Hubble modes with $k\eta\gtrsim 1$ contribute.

If the background is unpolarized, the occupation number satisfies $n_\lambda(k) = n(k)$. In this case,
\begin{equation}
    \rho_{\rm GW}(k) = \frac{1}{\pi^2a^4} k^4n(k).
\end{equation}
The energy density of the stochastic gravitational wave background is often characterized in a dimensionless way by
\begin{align}
    \Omega_{\rm GW}(f) = \frac{8\pi G}{3H_{0}^2}\rho_{\rm GW}(f)
\end{align}
with $H_0 = 100 h_0~{\rm km/s/Mpc}$ the Hubble scale today and $h_0 \approx 0.7$ but it was left unspecified in the rest of the paper. The frequency $f$ is given by $2\pi f = k$.

\subsection{Epoch Transitions}
An interesting application of the formulas in this section is the computation of amplification of the inflationary primordial spectrum. For simplicity, we work in a toy model of cosmology with only three epochs, de Sitter inflation, radiation-dominated, and matter-dominated epochs, and the transitions are assumed to be instantaneous. In this simplified cosmology, the scale factor reads
\begin{align}\label{eq:cosmo_a}
a(\eta)=\left\{\begin{array}{lll}
-\frac{1}{H_{\rm dS}\eta}, &\quad -\infty<\eta<\eta_{\rm r} & \text { (dS) } \\
&\\
\frac{1}{H_{\rm dS}\eta_{\rm r}^2}(\eta-2\eta_{\rm r}), &\quad \eta_{\rm r}<\eta<\eta_{\rm eq} &\text { (RD) } \\
&\\
a_{\rm eq}\frac{(\eta_{\rm eq} - 4\eta_{\rm r} +\eta)^2}{4(\eta_{\rm eq}-2\eta_{\rm r})^2}, &\quad \eta_{\rm eq}<\eta &  \text { (MD) }
\end{array}\right.
\end{align}
At each junction, the Hubble scale and its first derivative are continuous. In the following calculations, we assume $a_{\rm eq} \approx 1/3400$ and $H_{\rm dS} \lesssim 2.5\times 10^{-5}~M_{\rm Pl}$ \cite{Planck:2018jri}. Choosing the Bunch-Davies vacuum as an initial state, we have
\begin{equation}\label{eq:BD_vacuum}
    \xi_k(\eta) = \left(1-\frac{i}{k\eta}\right)e^{-ik\eta}.
\end{equation}
Moreover, from eq.~(\ref{eq:x_a}) , the general solution for the Bogoliubov coefficients is
\begin{subequations}
\begin{align}
        \alpha(\eta) &= c_1\left(1- \frac{i}{k\eta}- \frac{1}{2k^2\eta^2}\right)+c_2 \left(\frac{e^{2ik\eta}}{2k^2\eta^2}\right), \\
        \beta(\eta) & = c_1 \left(\frac{e^{-2ik\eta}}{2k^2\eta^2}\right) + c_2 \left(1+ \frac{i}{k\eta}- \frac{1}{2k^2\eta^2}\right),
\end{align}
\end{subequations}
where $c_1$ and $c_2$ are fixed by the Bunch-Davies vacuum, arbitrarily deep into the inflationary regime: $c_1 \to 1$, $c_2 \to 0$. This choice of vacuum implies that $N_\lambda =0$ in the formulas of the last section, such that $n_\lambda= |\beta|^2 $.

As can be seen from \eqref{eq:x_a} $\alpha$ and $\beta$ remain constant during RD, thus we can evaluate them at the beginning of that phase, $\eta_{\rm r}$, and denote them as
\begin{subequations}\label{eq:flat_beta}
\begin{align}
    \alpha_{\rm r} &= 1- \frac{i}{k\eta_{\rm r}}- \frac{1}{2k^2\eta_{\rm r}^2},  \\
    \beta_{\rm r} &= \frac{1}{2k^2 \eta_{\rm r}^2}e^{-2ik\eta_{\rm r}},
\end{align}
\end{subequations}
during RD. Assuming an instantaneous radiation- to matter-domination transition and eq.~(\ref{eq:flat_beta}) as initial conditions at the beginning of MD, after $\eta_{\rm eq}$ we have
\begin{subequations}
    \begin{align}
        \alpha(\eta) &= d_1 \left(1- \frac{i}{k\tau}- \frac{1}{2k^2 \tau^2}\right)+d_2\left(\frac{e^{2ik\eta}}{2k^2\tau^2}\right), \\
        \beta(\eta) &= d_1\left(\frac{e^{-2ik\eta}}{2k^2 \tau^2}\right) + d_2\left(1 + \frac{i}{k\tau}- \frac{1}{2k^2 \tau^2}\right),
    \end{align}
\end{subequations}
where we define $\tau \equiv \eta + \eta_{\rm eq} - 4\eta_{\rm r}$ and the coefficients  
\begin{subequations}
    \begin{align}
        d_1 &= \alpha_{\rm r}\left(1+ \frac{i}{k\tau_{\rm eq}}- \frac{1}{2k^2 \tau^2_{\rm eq}}\right)-\beta_{\rm r}\left(\frac{e^{2ik\eta_{\rm eq}}}{2k^2\tau^2_{\rm eq}}\right), \label{eq:d1} \\
        d_2 &= \beta_{\rm r}\left(1 - \frac{i}{k\tau_{\rm eq}}- \frac{1}{2k^2 \tau^2_{\rm eq}}\right) - \alpha_{\rm r}\left( \frac{e^{-2ik\eta_{\rm eq}}}{2k^2 \tau^2_{\rm eq}}\right), \label{eq:d2}
    \end{align}
\end{subequations}
with $\tau_{\rm eq} \approx 2\eta_{\rm eq}$.

Although the spectrum is affected by vacuum amplification during matter domination, this will only be effective for $k\eta_{\rm eq}\ll 1$, because the sub-Hubble modes with $k\eta_{\rm eq}>1$ at matter-radiation equality are still sub-Hubble ($k\eta_0\gg 1$) today and were never amplified. Therefore, we can approximate the spectrum in the sub-Hubble and super-Hubble limits
\begin{align}\label{eq:GR_power_spec}
    \rho_{\rm GW}(k) \simeq \begin{cases}
    \frac{9}{64\pi^2}\frac{1}{k^2\eta_{\rm eq}^2\eta_{\rm r}^4}, \quad k\eta_{\rm eq} < 1 \\
    \\
    \frac{1}{4\pi^2}\frac{1}{\eta_{\rm r}^4},\quad k\eta_{\rm eq}>1.
    \end{cases}
\end{align}
The primordial spectrum is scale-invariant up to the extra amplification during MD. We see that there is a suppression factor proportional to $k^{-2}$ compared to the spectrum before the matter-radiation equality. This is the typical behavior of the transfer function for gravitational waves \cite{Caprini:2018mtu}. Note that $\eta_{\rm r}$, and hence $\rho_{\rm GW}(k)$ can be written in terms of $H_{\rm dS}$ and known cosmological parameters as
\begin{equation}
    \eta_{\rm r}^4 =\frac{(1+z_{\rm eq})^4}{H^2_{\rm eq}H^2_{\rm dS}} = \frac{\Omega_{\rm r,0}^3}{2H_0^2 \Omega_{\rm m,0}^4}\frac{(1+z_{\rm eq})^4}{H_{\rm dS}^2},
\end{equation}
but in the rest of the paper the results are left in terms of $\eta_{\rm r}$. The associated frequency scale is
\begin{equation}
    f_{\rm r} = \frac{1}{2\pi |\eta_{\rm r}|} \sim 10^8\left(\frac{H_{\rm dS}}{10^{-4} \text{M}_{\rm Pl}}\right)^{1/2} \text{Hz}.
\end{equation}
Moreover,
\begin{equation}
    f_{\rm eq} = \frac{1}{2\pi \eta_{\rm eq}} \sim 10^{-16} \text{Hz}
\end{equation}
is an upper bound on the frequency of modes which are amplified during matter domination.

In this section have assumed instantaneous transitions between different epochs and neglected the recent phase of cosmic acceleration. For a continuous numerical integration of eqs.~(\ref{eq:x_a}) and (\ref{eq:y_a}), see \cite{Mendes:1994ai}, and for more discussion about the cosmic evolution of the primordial power spectrum, see \cite{Caprini:2018mtu}.

\section{Vacuum Amplification in Chern-Simons Gravity}\label{sec:vacuum_amp_CS}
\subsection{CS modified Gravitational Waves}
The 4D action of CS gravity is given by
\begin{align}
    \mathcal{S} = \mathcal{S}_\mathrm{EH} + \mathcal{S}_\mathrm{CS} + \mathcal{S}_\varphi.
\end{align}
The Einstein-Hilbert action in GR 
\begin{align}
    \mathcal{S}_\mathrm{EH} &= \frac{1}{2\kappa^2}\int d^4 x \sqrt{-g} R
\end{align}
is modified by additional parity-violating CS term and the pseudo-scalar dynamics terms
\begin{align}
    \mathcal{S}_\mathrm{CS} &= \frac{\alpha}{4} \int d^4 x \sqrt{-g}~ \varphi {}^*RR\\
    S_{\varphi} &= -\beta \int d^4 x \sqrt{-g} \left[\frac{1}{2}\left(\nabla^\mu \varphi\right)\left(\nabla_\mu \varphi\right)+V(\varphi)\right],
\end{align}
where $\kappa = M_{\rm Pl}^{-1} = \sqrt{8\pi G}$, $\alpha$ has mass dimension equal to minus one, and $\beta$ is a dimensionless constant equal to one or zero for dynamical and non-dynamical CS gravity, respectively. The Chern-Simons coupling constant $\alpha$ is often written as
\begin{align}
    \alpha = \frac{\ell_{\rm CS}^2}{2\kappa},
\end{align}
with the CS characteristic lengthscale, $\ell_{\rm CS} \lesssim 10^8~\mathrm{km}$, constrained by measurements of frame-dragging effects around the Earth\footnote{The more stringent constraint $\ell_{\rm CS} \lesssim 8.5$ km was found in \cite{Silva:2020acr}, after combining observations of the GW profile of neutron stars mergers and X-ray emission from an isolated neutron star.} \cite{Ali-Haimoud:2011zme}. The quantity 
\begin{align}
    {}^*RR =\frac{1}{2} \varepsilon^{a b e f} R_{a b c d} R^{c d}{ }_{e f}
\end{align}
is known as the Prontryagin density with $R_{abcd}$ the Riemann curvature tensor and $\varepsilon_{abcd}$ the Levi-Civita tensor. 

In order to get the equation of motion for GWs, we expand the action to the second order of metric perturbation (and assume the TT gauge) \cite{Alexander:2004wk},
\begin{align}
    \begin{aligned}
    \mathcal{S}^{(2)}_\mathrm{GW} &= \frac{1}{8\kappa^2} \int d^4 x \Big(a^2(\eta)  \left[(h_{ij}^\prime)^2 - (\partial_k h_{ij})^2 \right] \\
    &- 2\kappa^2 \alpha\varphi'\varepsilon^{ijk}\Big[(h_{~i}^q)'(\partial_j h_{kq})' -\left(\partial^r h^q{ }_i\right) \partial_j \partial_r h_{k q}\Big]\Big),
    \end{aligned}
\end{align}
where $\varepsilon^{i j k} \equiv \varepsilon^{0 i j k}$. In Fourier space, using circular polarization basis
\begin{subequations}\label{circular_pol}
\begin{align}
    e_{ij}^{\rm R}(\hat{k}) &= \frac{1}{\sqrt{2}}\left(e_{ij}^{+}(\hat{k}) + ie_{ij}^{\times}(\hat{k})\right), \\ 
    e_{ij}^{\rm L}(\hat{k}) &= \frac{1}{\sqrt{2}}\left(e_{ij}^{+}(\hat{k}) - ie_{ij}^{\times}(\hat{k})\right),
\end{align}
\end{subequations}
we can decompose the metric perturbations as
\begin{align}
    \begin{aligned}
        h_{ij}(\eta,\mathbf{x})  &= \sqrt{16\pi G}\sum_{\lambda= R,L} \int \frac{d^3 k}{(2 \pi)^3}\frac{\mathcal{Q}_\lambda(\eta, \mathbf{k})}{\sqrt{2k}} e_{ij}^\lambda(\hat{k})e^{i\mathbf{k}\cdot \mathbf{x}}, \\
        \mathcal{Q}_\lambda(\eta, \mathbf{k})&= \hat{a}_\lambda(\mathbf{k}) \frac{\mu_\lambda(\eta, \mathbf{k})}{z_\lambda(\eta, k)} +\hat{a}_\lambda^{\dagger}(-\mathbf{k}) \frac{\mu_\lambda^*(\eta, -\mathbf{k})}{z_\lambda(\eta,k)},
    \end{aligned}
\end{align}
where
\begin{align}
    z_\lambda(\eta, k) \equiv a(\eta)\sqrt{1- 2\kappa^2 \alpha\lambda_{R,L} k \frac{\dot{\varphi}}{a}},
    \label{eq:cs_scale_factor}
\end{align}
with $\lambda_{\rm R} = 1$, $\lambda_{\rm L} = -1$. The circular polarization tensors \eqref{circular_pol} satisfy
\begin{subequations}
\begin{align}
    \epsilon_m^{\;\;\;np}\hat{k}_pe^\lambda_{ln}(\hat{k}) &= i \lambda e^{\lambda}_{lm}(\hat{k}), \\ e^{\lambda}_{mn} (-\hat{k}) e_{\tilde{\lambda}}^{mn} (\hat{k}) &=  2\delta^{\lambda}_{\tilde{\lambda}}.
\end{align}
\end{subequations}
The mode function $\mu_\lambda(\eta,{\mathbf{k}}) =\mu^\lambda_k(\eta)$ satisfy \cite{Choi:1999zy, Alexander:2004wk}
\begin{align}
    (\mu_k^\lambda)'' + \left(k^2 - \frac{(z_{k}^\lambda)''}{z_{k}^\lambda}\right)\mu_{k}^\lambda = 0 \label{eq:CS-eom}.
\end{align}
Compared to GR, the dCS coupling introduces a new physical scale, defined by 
\begin{equation}\label{kCS}
    k_{\rm CS} = \frac{1}{\alpha \kappa^2 |\dot{\varphi}|}.
\end{equation}
As we shall see, the dCS contribution to the power spectrum is proportional to $k/k_{\rm CS}$, and so the GR spectrum is significantly modified on scales such that $k> k_{\rm CS}$.

\subsection{Bogoliubov coefficients and energy power spectrum in Chern-Simons gravity}\label{sec:bog_coeff_dCS}

From equation \eqref{eq:CS-eom} for the gravitational wave mode function, we see that vacuum amplification in dCS can easily be described after extending the formalism of Sec.~\ref{sec:vacuum_amp}. The chords correspond now to periods of constant $z'_\lambda$ instead of $a'$. Another difference is the $k$ dependence of $z_\lambda$ which does not affect analysis for the continuous-time evolution of the Bogoliubov coefficients.  More explicitly, we can calculate the particle production in dCS after solving 
\begin{subequations}
    \begin{align}
        & \alpha_\lambda^{\prime}(\eta)=\frac{i}{2 k}\left[\alpha_\lambda(\eta)+\beta_\lambda(\eta) e^{2 i k\left(\eta-\tilde{\eta}\right)}\right] \frac{(z^\lambda_k)^{\prime \prime}(\eta)}{z^\lambda_k(\eta)} \label{eq:alpha_dCS_prime}, \\
        & \beta_\lambda^{\prime}(\eta)=-\frac{i}{2 k}\left[\beta_\lambda(\eta)+\alpha_\lambda(\eta) e^{-2 i k\left(\eta-\tilde{\eta}\right)}\right] \frac{(z^\lambda_k)^{\prime \prime}(\eta)}{z^\lambda_k(\eta)}, 
        \label{eq:beta_dCS_prime}
    \end{align}
\end{subequations}
or, using \eqref{eq:ab_to_XY},
\begin{align}
& X_\lambda^{\prime \prime}+\left(k^2-\frac{(z_{k}^\lambda)''}{z_{k}^\lambda}\right) X_\lambda =0, \label{eq:CS-x_a}\\
& Y_\lambda= \frac{i}{k} X_\lambda' \label{eq:CS-y_a}.
\end{align}
However, in the dCS case the evolution equation depends on the polarization and hence the vacuum amplification is parity-violating in general. 

In GR, vacuum amplification occurs due to the time-dependence part of the effective frequency in eq.~(\ref{eq:x_a}), i.e., $a''/a$. Modes with wavenumber $k$ smaller than $a''/a$ are amplified. From \eqref{eq:x_a}, in dCS gravity the time-dependent part of the effective frequency is 
\begin{equation}\label{eq:ddz_z}
    \frac{z''}{z} \simeq \frac{a''}{a}+ \frac{a''}{a}\lambda_{R,L}\alpha\kappa^2k\frac{\dot{\varphi}}{a}-\lambda_{R,L}\alpha\kappa^2k\left(\frac{a'}{a} \Ddot{\varphi} +a\dddot{\varphi}\right),
\end{equation}
from which we see that the time-dependent part of the effective frequency now depends also on $k_{\rm CS}$ and its time derivatives. Thus, not only vacuum amplification is affected by the dCS coupling but also the dCS pseudo-scalar appears as a potential source for the amplification. Even for constant scale factor, or if the time scale $\Delta\eta_{\rm CS}$ associated to changes in $\dot{\varphi}$ is much smaller than a Hubble time, modes with $k\Delta\eta_{\rm CS}< 1$ can be amplified. 

The energy density and power spectrum of gravitational waves in dCS gravity receive a contribution from the pseudo-scalar Pontryagin coupling. This can be computed from the energy-momentum tensor for gravitational waves in the dCS theory. For $\nabla_\mu \varphi = \delta_\mu^0 \dot{\varphi}$, where the dot denotes derivative with respect to cosmic time, we have \cite{Isi:2018miq}
\begin{equation}
    T_{\mu\nu} = T_{\mu\nu}^{(\rm GR)} + \frac{\alpha \dot{\varphi}}{2}\varepsilon_{m}^{\;\;\;np}\langle\langle \nabla_{(\mu}h^{m\beta}\nabla_{\nu)}\nabla_p h_{\beta n}\rangle\rangle, 
\end{equation}
where $T^{(\rm GR)}_{\mu\nu}$ is given by eq.~(\ref{eq:stress_tensor_gr}). The new contribution coming from the non-minimal coupling of the dCS scalar to the energy density is then
\begin{align}
    \rho_{\rm GW}^{(\rm CS)} &= \frac{\alpha\kappa^2 \dot{\varphi}}{a^{2}}\varepsilon_{m}^{\;\;\;np}\sum_{\lambda, \tilde{\lambda}}\int \frac{d^3 \mathbf{k}}{(2\pi)^3}\frac{d^3 \mathbf{\tilde{k}}}{(2\pi)^3}\frac{e^{\lambda mq}(\hat{k})e^{\tilde{\lambda}}_{qn}(\hat{\tilde{k}})}{2\sqrt{k \tilde{k}}} \times\nonumber\\
    &\times (i\tilde{k}_p) \langle  \mathcal{Q}_\lambda^{\dagger\prime} (-\mathbf{k}, \eta) \mathcal{Q}_{\tilde{\lambda}}^{\prime} (\mathbf{\tilde{k}},\eta)\rangle e^{i(\mathbf{k}+ \mathbf{\tilde{k}})\cdot \mathbf{x}}.
\end{align}
For the dCS case, $\mathcal{Q}_\lambda'$ includes terms proportional to $\alpha \dot{\varphi}$, but since we are interested in $\rho_{\rm GW}^{(\rm CS)}$ to the leading order in $\alpha$, we can neglect such terms. Hence, for sub-Hubble modes, we get
\begin{align}
    \rho_{\rm GW}^{(\rm CS)} &= -\frac{\alpha \kappa^2\dot{\varphi}}{2\pi^2 a^{4}(\eta)}\sum_\lambda \lambda_{\rm R,L} \int dk k^4\left(1+2|\beta_\lambda|^2\right)\times \nonumber\\
    &\times \left(N(\mathbf{k})+ N(-\mathbf{k})+1\right),
\end{align}
where we only included the $\alpha$-independent part of $z_\lambda(k, \eta) = a(\eta) + \mathcal{O}(\alpha)$ in the denominator. 
The total energy density is then
\begin{align}\label{eq:dCS_rho}
    \rho_{\rm GW} = \frac{1}{4\pi^2 a^4(\eta)}\sum_\lambda \int dk \left[1- 2\lambda_{\rm R,L}\left(\frac{k}{k_{\rm CS}}\right)\right]k^3 \nonumber\\
    \times\left(1+2|\beta_\lambda|^2\right)\left(2N(k)+1\right)
\end{align}
for an initially isotropic spectrum and with $k_{\rm CS}$ given by \eqref{kCS}. 

The energy density spectrum is now
\begin{align}\label{eq:dCS_rho_k}
    \rho_{\rm GW}(k) = \frac{d\rho_{\rm GW}}{d\ln k}= \frac{1}{4\pi^2 a^4(\eta)}\sum_\lambda \left[1- 2\lambda_{\rm R,L}\left(\frac{k}{k_{\rm CS}}\right)\right]k^4 \nonumber\\
    \times\left(1+2|\beta_\lambda|^2\right)\left(2N(k)+1\right).
\end{align}
Since $\lambda_{\rm R,L} = \pm 1$, this spectrum will generically be polarized. Note that the leading-order difference from the GR result also includes a contribution from $|\beta_\lambda|^2$, which satisfies equation \eqref{eq:beta_dCS_prime} and hence includes terms $\mathcal{O}(\alpha)$, which multiply the GR contribution in the formula above. Explicitly, expanding $\beta_\lambda$ as $\beta_\lambda = \beta^{(0)}_\lambda + \beta_\lambda^{(1)}$, where the superscripts denote the order dependence on the dCS coupling $\alpha$, we have
\begin{equation}\label{eq:total_bsquared}
    |\beta_\lambda|^2 = |\beta^{(0)}_\lambda|^2 + 2\text{Re}\left(\beta^{(0)}_\lambda \beta^{(1)*}_\lambda\right) +\mathcal{O}\left(\left(\alpha\kappa^2 \dot{\varphi}k\right)^2\right),
\end{equation}
and the second term will produce an $\mathcal{\alpha}$ contribution to $\rho_{\rm GW}(k)$.

\section{Applications}\label{sec:application}
In this section, we explore the vacuum amplification in Chern-Simons gravity explicitly for four different settings. The general rationale is that $\varphi$ has a potential that dictates its homogeneous evolution, and such an evolution modifies the vacuum of the tensor fluctuations. Since $R\tilde{R}$ vanishes for the FLRW metric\footnote{The vanishing of the Pontryagin term for the FLRW metric is an identity due to the symmetries of the metric. This can be seen after explicitly evaluating the contraction $R\tilde{R}$ for that metric, which yields zero. Another way to see this is through the fact that $R\tilde{R} = C\tilde{C}$ where $C$ represents the Weyl tensor \cite{Grumiller:2007rv}. So, the Pontryagin term vanishes for any conformally flat metric, which is the case of the FLRW metric.}, at the background level the metric dependence on evolution of $\varphi$' is established by the minimal coupling through the Klein-Gordon equation in curved spaces. 

Although the general formalism established in the previous section works for any initial state, in this section we consider an initial flat power spectrum from (dS) inflation. Our goal is to compute the leading CS corrections ($\mathcal{O}(k/k_{\rm CS})$) to the Bogoliubov coefficients and the energy density spectrum in the following cases: for flat space with a mass potential; for constant $\dot{\varphi}$, i.e. the nondynamical CS gravity; for a continuous evolution parametrized by an effective fluid equation of state $w_\varphi$; and for transitions between different periods of constant-$\dot{\varphi}$ evolution, which we call dCS transitions. 

As we shall see, for certain values of $k$, the energy power spectrum is polarized and so we also calculated the chirality parameter
\begin{align}
    \Delta\chi(k) = \frac{\rho_{\rm GW}^{\rm(R)}(k)- \rho_{\rm GW}^{\rm(L)}(k)}{\rho_{\rm GW}^{\rm(R)} (k)+ \rho_{\rm GW}^{\rm(L)}(k)}.
\end{align}
for all relevant cases.

\subsection*{Case I: Minkowski Limit}

To show the universality of the CS modification to the vacuum amplification, we first consider the Minkowski limit case, where we set $a(\eta) \to 1$. This limit also approximates the realistic case of $\varphi$ evolution happening on a time scale much smaller than the cosmological expansion scale. 

Consider the case where the scalar field $\varphi$ has a potential 
\begin{equation}
    V(\varphi) = \frac{1}{2}m^2\varphi^2,
\end{equation}
with $m$ being the mass of the field. 
To solve the equation of motion for $\varphi$, we start from  
\begin{equation}
\dot{\varphi} = \sqrt{2(\rho_\varphi-V)}
\end{equation}
where the energy density $\rho_\varphi$ is constant due to the  $\varphi$'s equation of motion. Thus, 
\begin{equation}
\begin{aligned}
    t &= \int^{\varphi}_{\varphi_0} \frac{\mathrm{d} \varphi}{\sqrt{2(\rho_\varphi-V)}} = \frac{1}{m} \arcsin \left(\frac{m}{\sqrt{2\rho_\varphi}}( \varphi- \varphi_0)\right),
\end{aligned}
\end{equation}
from which we obtain $\varphi(t)$ as 
\begin{equation}
\varphi(t) = \frac{\sqrt{2\rho_\varphi}}{m}\sin{[m(t-t_0)]}.
\end{equation}
From (\ref{eq:ddz_z}), the time-dependent contribution for the effective frequency of the graviton mode function is, to leading order in $k/k_{\rm CS}$, 
\begin{equation}
\frac{z_\lambda''}{z_\lambda} = \lambda_{\rm R, L} \alpha\kappa^2 k m^2 \sqrt{2\rho_\varphi} \cos{[m(t-t_0)]}
\end{equation}
Plugging this back into eq.~(\ref{eq:x_a}) gives the Mathieu equation
\begin{equation}\label{eq:mathieu}
X_{\lambda}^{\prime \prime}+\left[k^2 + M_0 k \cos{(m\Delta t)}\right] X_\lambda=0,  
\end{equation}
where $M_0 = -\lambda_{\rm R, L} \alpha\kappa^2 m^2 \sqrt{2\rho_\varphi}$ and $ \Delta t= t-t_0$. This implies a parametric resonance for $X$. 

Following standard Floquet analysis \cite{landau2013course,Traschen:1990sw}, we treat the last term of the above equation as a time-dependent oscillatory perturbation. It is known that for parametric resonance, the instability bands are centered at 
\begin{equation}
k^{(n)}=\frac{n}{2} m,
\end{equation}
outside of which the solution of the unperturbed equation is very stable, which means no particle production. However, inside the instability bands, we can analyze the solution perturbatively. Consider the first band with $n=1$, we take 
\begin{align}
    k = \frac{m}{2} + \delta_k \left(\frac{M_0k}{2m}\right),
\end{align}
with $\delta_k \in [-1, 1]$. Here, the width of the band is linearly dependent of the perturbation amplitude. For solutions inside this band, we take the following ansatz:
\begin{equation}
X_{\lambda}(t)=A(t) \cos\left( \frac{m}{2} \Delta t\right)+B(t) \sin \left(\frac{m}{2} \Delta t\right),
\end{equation}
where $A(t)$ and $B(t)$ are coefficient functions. Inserting this back to our Mathieu equation, we have \cite{Traschen:1990sw} 
\begin{subequations}
\begin{align}
& \ddot{A}+m \dot{B}-\left(\frac{m}{2}\right)^2 A=-k^2 A-\frac{M_0 k}{2} A \\
& \ddot{B}-m \dot{A}-\left(\frac{m}{2}\right)^2 B=-k^2 B-\frac{M_0 k}{2} B,
\end{align}
\end{subequations}
where higher frequency terms were dropped.
Ignoring the higher order $\ddot{A}$ and $\ddot{B}$, and assuming 
\begin{equation}
A(t) \sim e^{s \Delta t} \quad\text { and }\quad B(t) \sim e^{-s \Delta t} ,
\end{equation}
we have 
\begin{equation}
s= \pm s_k =\pm \frac{M_0 k}{2 m}\left( 1 + \delta_k \right)^{1 / 2} 
\end{equation}
Since the polarization is encoded in $s_k$, we can only consider $+s_k$. Thus, we have the general solution for eq.~(\ref{eq:mathieu}) 
\begin{equation}
X_{\lambda}= C e^{s_k \Delta t} \cos\left(\frac{m}{2} \Delta t\right) +D e^{-s_k \Delta t} \sin\left(\frac{m}{2} \Delta t\right). 
\end{equation}
Using the initial condition where Bogoliubov coefficients satisfy $\alpha_{\lambda}(t=t_0)=1$ and $\beta_{\lambda}(t=t_0)=0$, we find
\begin{equation}
X_{\lambda}(t)= e^{s_k \Delta t} \cos\left(\frac{m}{2} \Delta t\right) - \frac{2(ik+s_k)}{m} e^{-s_k \Delta t} \sin\left(\frac{m}{2} \Delta t\right). 
\end{equation}
The coefficient $\beta_{\lambda}(t)$ is found from eq.~(\ref{eq:beta_XY}) and the resulting number density of gravitons is given by 
\begin{widetext}
\begin{align}\label{eq:n_minkowski}
n_\lambda (t) = |\beta_{\lambda}(t)|^2= \frac{1}{4} & \Bigg\{ \left(e^{s_k \Delta t}- e^{-s_k \Delta t}\right)^2\left(1+\frac{s_k^2}{k^2}\right)\cos^2\left(\frac{m}{2} \Delta t\right) + \frac{m^2}{4k^2}\left[e^{s_k \Delta t}-\frac{4k^2}{m^2} \left(1+\frac{s_k^2}{k^2}\right)e^{-s_k \Delta t} \right]^2 \sin^2\left(\frac{m}{2} \Delta t\right)  \nonumber\\
&- \frac{s_k m}{k^2}\left(e^{s_k \Delta t}- e^{-s_k \Delta t} \right)\left[e^{s_k \Delta t}-\frac{4k^2}{m^2} \left(1+\frac{s_k^2}{k^2}\right)e^{-s_k \Delta t}\right]\cos\left(\frac{m}{2} \Delta t\right)\sin\left(\frac{m}{2} \Delta t\right) \Bigg\}.
\end{align}
\end{widetext}
The energy density of produced gravitational waves can be obtained by eq.~(\ref{eq:dCS_rho}).  Terms that are proportional to $e^{2s_k\Delta t} + e^{-2s_k\Delta t}$ are parity-even and the parity-odd part of the number density is suppressed by factors of $|M_0|/m$. The latter holds because the frequency at the center of the band is set by $m$, while $|M_0| \sim \alpha \kappa^2 m^2 |\dot{\varphi}| \sim m^2/k_{\rm CS}$, such that $|M_0|/m \sim k/k_{\rm CS}$. Even for a background pseudo-scalar field with $\rho_\varphi \lesssim \rho_c$ and the induced GWs with $f < \mathcal{O}(10^4)~\mathrm{Hz}$, the parity-violation number density is proportional to $(M_0/m) \lesssim \mathcal{O}(10^{-7})$.
\begin{figure}[h]
    \centering
    \includegraphics[width=1.1\linewidth]{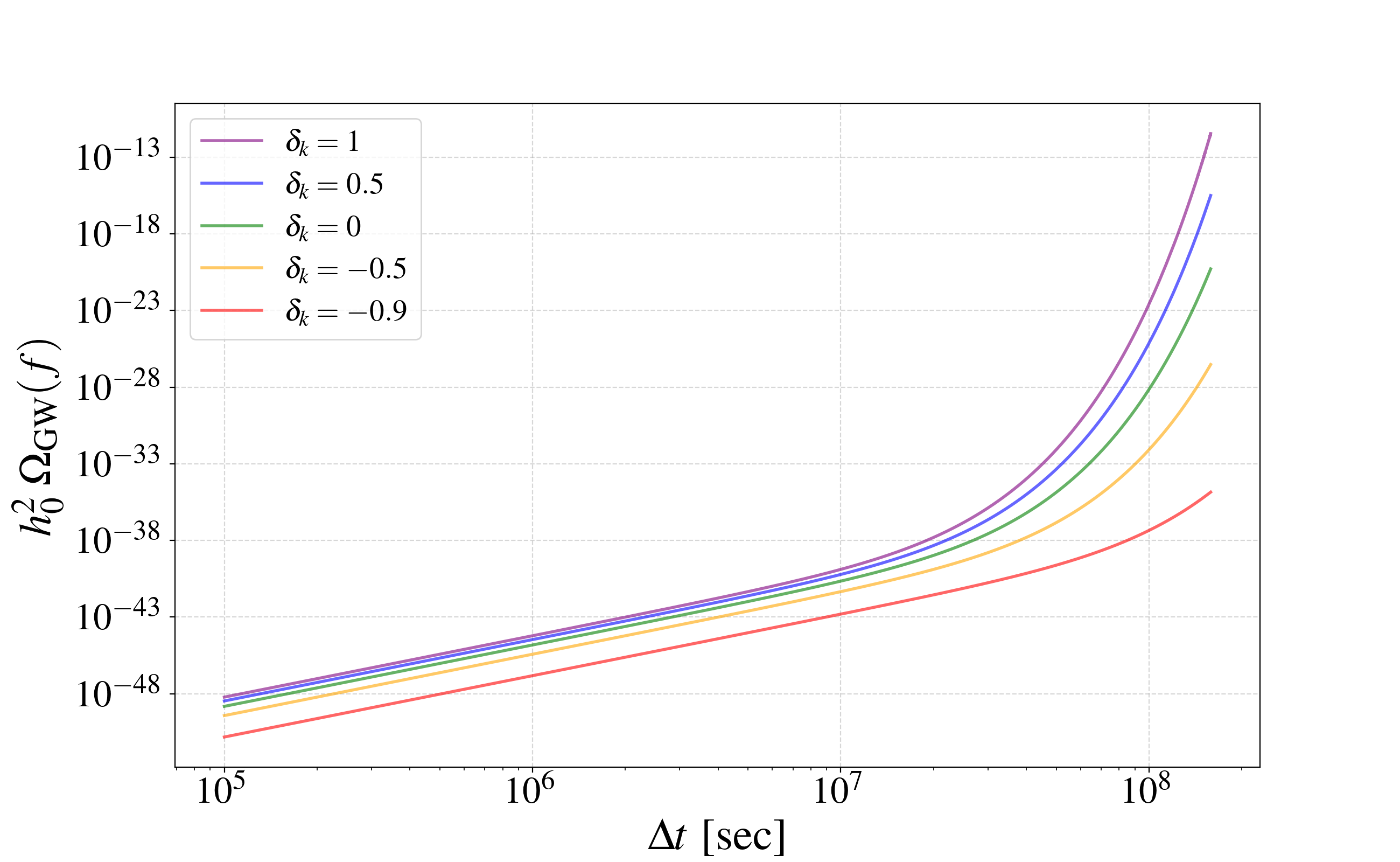}
    \caption{The dCS vacuum amplification in the Minkowski limit. We assume the CS pseudo-scalar has mass $m = 10^{-12}~\mathrm{eV}$, corresponding to $f \sim \mathcal{O}( 10^2)~\mathrm{Hz}$ within the frequency band of LIGO-Virgo-KAGRA. The CS characteristic length is set to $\ell_{\rm CS} = 10^8~\mathrm{km}$, and the pseudo-scalar energy density $\rho_\varphi$ is taken as $10\%$ of the critical energy density $\rho_{\rm c}$. Since the power spectrum is nearly parity-even, we only show the right-handed modes here. Within the instability band, parametric resonance results in enhanced particle production at higher frequencies, while graviton production diminishes as $\delta_k$ decreases. As $\delta_k \to -1$, particle production converges to zero. }
    \label{fig:minkowsk}
\end{figure} 

Thus, for a cosmological $\varphi$ evolution, the amount of amplification induced by the dCS coupling is negligible and the energy power spectrum is dominated by the GR contribution and hence even (see Fig. \ref{fig:minkowsk}). However, for an astrophysical source of stochastic GWs, the bound on parity-violation can be relaxed by setting $\rho_\varphi$ to be much larger than the critical energy density $\rho_c$. One can consider gravitational wave induced by a dark matter halo made of axion-like particles with a Chern-Simons coupling to gravity. The energy density of the dark matter halo is large enough to have observable chiral gravitational waves \cite{Yoshida:2017cjl}.

\subsection*{Case II: Constant $\dot\varphi$}\label{sec:constant_phidot}
Assuming a constant $\dot{\varphi}$ we can compute $\beta_\lambda$ after solving eqs.~(\ref{eq:x_a}) and (\ref{eq:y_a}). A bound on the non-dynamical Chern-Simons profile $|\dot\varphi| \lesssim 0.4~ \mathrm{km}$ is set by binary pulsar systems \cite{Ali-Haimoud:2011wpu}. The corresponding frequency is $f_{\rm CS} \gtrsim 10^5~{\rm Hz}$. The Chern-Simons correction factor can be perturbatively expanded into\footnote{The more stringent bound of \cite{Yunes:2008ua} correspond to a larger lower bound for $f_{\rm CS}$, which improves the $f/f_{\rm CS}$ perturbative analysis.} 
\begin{align}
    z_\lambda(\eta, f) \approx a - \lambda_{\rm R,L}(f/f_{\rm CS})
\end{align}
if we restrict
\begin{align}\label{eq:cutoff-freq}
    f \ll \frac{1}{2}f_{\rm CS}\left(\frac{T_0}{\Lambda_{\rm CS}}\right),
\end{align}
where $\Lambda_{\rm CS}$ is the cut-off scale for dCS gravity and $T_0$ is the temperature of the cosmic microwave background radiation today. For modes that violate this bound, there is a tachyonic instability in the mode equation from early times on. 

Since we are interested in the leading-order modification to the GR result, we solve the Bogoliubov coefficients perturbatively. Note that, for constant $\dot{\varphi}$, we have
\begin{equation}
    \frac{z''_\lambda}{z_\lambda} \simeq \frac{a''}{a}\left(1+ \frac{\lambda_{\rm R,L}}{a}\left(\frac{k}{k_{\rm CS}}\right)\right), 
\end{equation}
and so there is $X_\lambda$ and the Bogoliubov coefficients are not modified during radiation domination. Assuming the initial flat spectrum in eq.~(\ref{eq:GR_power_spec}), we then need to solve eq.~(\ref{eq:x_a}) during matter domination. 

Since we are interested in the leading order dCS contribution, we write $X_\lambda = X_\lambda^{(0)} + X_\lambda^{(1)}$, with $X_\lambda^{(1)}$ proportional to $\alpha \dot{\varphi}$. Plugging this decomposition into eq.~(\ref{eq:x_a}) and matching terms of the same order gives
\begin{align}\label{eq:const_phidot_eom}
    X_\lambda^{(0)''} + \left(k^2 - \frac{2}{\tau^2}\right)X_\lambda^{(0)} &= 0, \\
    X_\lambda^{(1)''} + \left(k^2 - \frac{2}{\tau^2}\right)X_\lambda^{(1)} &= J_\lambda(\tau, k), 
\end{align}
where $\tau \approx \eta + \eta_{\rm eq}$ and the source term is 
\begin{align}
    J_\lambda(\tau, k) = \frac{\lambda_{\rm R,L}}{a}\left(\frac{k}{k_{\rm CS}}\right)\frac{a''}{a}X_\lambda^{(0)}.
\end{align}

The solution for $X_\lambda^{(1)}$ has the form 
\begin{equation}\label{eq:X1_sol}
    X_\lambda^{(1)} = \int d\tilde{\tau}\; G(\tau, \tilde{\tau}) J_\lambda(\tilde{\tau},k), 
\end{equation}
where $G(\tau, \tilde{\tau})$ is the Green's function for he operator in the left-hand side of eq.~(\ref{eq:const_phidot_eom}). Since the source term vanishes for $\tau<\tau_{\rm eq}$, we have $G(\tau, \tilde{\tau}) =0=G'(\tau, \tilde{\tau})$ for $\tau< \tilde{\tau}$. This fixes the Green's function to be
\begin{align}
    G(\tau, \tilde{\tau}) &= \Theta(\tau - \tilde{\tau})\Bigg[\frac{1}{k}\left(1+\frac{1}{k^2\tau\tilde{\tau}}\right)\sin k(\tau-\tilde{\tau}) \nonumber\\
    &\quad\quad\quad\quad +\frac{1}{k}\left(\frac{1}{k\tau}- \frac{1}{k\tilde{\tau}}\right)\cos k(\tau-\tilde{\tau})\Bigg],
\end{align}
with $\Theta(\tau -\tilde{\tau})$ the Heaviside function.

Since the integrand in \eqref{eq:X1_sol} only has support after matter-radiation equality, $X^{(0)}_\lambda$ has the form 
\begin{align}
    X^{(0)}_\lambda(\tau) &= \tilde{d}_1 \left(1-\frac{i}{k\tau}\right)e^{-ik\tau} + \tilde{d}_2 \left(1+ \frac{i}{k\tau}\right)e^{ik\tau},
\end{align}
Here we define $\tilde{d}_1 = d_1 e^{ik\eta_{\rm eq}}$ and $\tilde{d}_2 = d_2 e^{-ik\eta_{\rm eq}}$ so that all the $\eta$ dependence can be substituted with $\tau$. Hence,
\begin{align}
    \tilde{d}_1 &=  \alpha_{\rm r}\left(1+ \frac{i}{k\tau_{\rm eq}}- \frac{1}{2k^2 \tau^2_{\rm eq}}\right)e^{ik\eta_{\rm eq}}-\beta_{\rm r}\frac{e^{3ik\eta_{\rm eq}}}{2k^2\tau^2_{\rm eq}}, \\
    \tilde{d}_2 &= -\alpha_{\rm r} \frac{e^{-3ik\eta_{\rm eq}}}{2k^2 \tau^2_{\rm eq}} + \beta_{\rm r}\left(1 - \frac{i}{k\tau_{\rm eq}}- \frac{1}{2k^2 \tau^2_{\rm eq}}\right)e^{-ik\eta_{\rm eq}} .
\end{align}

The corrections to the Bogoliubov coefficient can be found from \eqref{eq:ab_to_XY} after evaluating the integral in \eqref{eq:X1_sol}. For sub-Hubble modes, the expression for $\beta^{(1)}_\lambda$ can be simplified by taking the limit $k\tau\to \infty$:
\begin{align}
    \beta^{(1)}_\lambda (k\eta) &\simeq -\lambda_{\rm R,L} \left(\frac{k}{k_{\rm CS}}\right)\left(\frac{4\eta_{\rm eq}^2 k^2}{a_{\rm eq}}\right)\mathcal{I}(x),\\
    \mathcal{I}(x) &\equiv \int_{k\tau_{\rm eq}}^{k\tau} \d\tilde{x}~\left(\frac{1+i\tilde{x}}{\tilde{x}^5}\right)X^{(0)}_\lambda(\tilde{x})e^{-i\tilde{x}}.
\end{align}
The integrand of $\mathcal{I}(k\eta)$ includes terms of the form $\tilde{x}^{-n}e^{-2i\tilde{x}}$, with $n>0$, and hence the result can be expressed in terms of the exponential integral function $\text{Ei}(\tilde{x})$:
\begin{widetext}
\begin{align}
    \mathcal{I}(k\tau) &= \left[\frac{i\tilde{d}_1}{15}\left(\frac{(2\tilde{x}^4 + i \tilde{x}^3- \tilde{x}^2 +6i \tilde{x}+3)}{\tilde{x}^5}e^{-2i\tilde{x}} + 4i\text{Ei}[-2i \tilde{x}] \right) -i\tilde{d}_2\left(\frac{1}{3\tilde{x}^3}+\frac{1}{5\tilde{x}^5}\right)\right]_{x_{\rm eq} = k\tau_{\rm eq}}^{x = k\tau} 
\end{align}
\end{widetext}
Since we are only interested in the modes inside the Hubble radius today, we can set $k\eta \gg 1$. Using the matter-radiation equality as a dividing line, we have 
\begin{align}\label{eq:b1_constdotphi}
    \beta_\lambda^{(1)} &=
    -\lambda_{\rm R,L} \left(\frac{k}{k_{\rm CS}}\right)\left(\frac{4\eta_{\rm eq}^2 k^2}{a_{\rm eq}}\right)
    \begin{cases}
        \frac{i\tilde{d}_2}{3(2k\eta_{\rm eq})^3} \;, \quad  k\eta_{\rm eq} >1\\
        \\
      \frac{i(\tilde{d}_2-\tilde{d}_1)}{5(2k\eta_{\rm eq})^5}\;. \quad k\eta_{\rm eq}<1
    \end{cases}
\end{align}
Or, taking the appropriate limits of the coefficients $\tilde{d}_1$ and $\tilde{d}_2$,
\begin{align}
    \beta_\lambda^{(1)} = 
    -\lambda_{\rm R,L} \left(\frac{k}{k_{\rm CS}}\right)\left(\frac{4\eta_{\rm eq}^2 k^2}{a_{\rm eq}}\right)
    \begin{cases}
        \frac{i\beta_{\rm r}}{3(2k\eta_{\rm eq})^3} \;, \quad  k\eta_{\rm eq} >1\\
        \\
       \frac{2(\alpha_{\rm r} + \beta_{\rm r})}{15(2k\eta_{\rm eq})^4}\;,  \quad k\eta_{\rm eq}<1
    \end{cases}
\end{align}
In both cases, the modification is negligibly small compared to the GR contribution. This happens because there is no particle production during the whole radiation dominated period and so the CS contribution to the effective frequency through $z_\lambda(\eta, k)$ is small when we get to matter domination. Quantitatively, $T_0/\Lambda_{\rm CS} \sim \mathcal{O}(10^{-12})$ for $\Lambda_{\rm CS} \sim 150$ MeV (close to the QCD scale) and $T_0/\Lambda_{\rm CS} \sim \mathcal{O}(10^{-4})$ for $\Lambda_{\rm CS} \sim 1$ eV (close to the recombination scale), and therefore the signal is very suppressed for perturbative frequencies ($k/k_{\rm CS} <T_0/ \Lambda_{\rm CS}$) even for $k\eta_{\rm eq}\ll 1$. Moreover, in non-dynamical CS gravity, the local and global evolution of $\varphi$ cannot be decoupled, and so the astrophysical bound on $k_{CS}$ does not allow us to take values of $k_{\rm CS}$ small enough for a significant contribution in the perturbative region $k<k_{\rm CS}$. However, the calculations in this section are relevant for the transitions considered in the next section. 

\subsection*{Case III: dCS transitions}

In this section, we start investigating the effect of a change in $\dot{\varphi}$ on the vacuum amplification mechanism. We are interested in the case where $\dot{\varphi}$ changes between two asymptotic behaviors. For analytical estimates, we assume a sharp transition as a proxy for the $\dot{\varphi}$ dynamics,
\begin{equation}\label{eq:two_stage_kcs}
    \frac{\alpha\lambda_{\rm R,L}\dot{\varphi}}{M_{\rm Pl}^2} = \lambda_{\rm R,L} \left[\frac{1}{k^{(1)}_{\rm CS}}+\left(\frac{1}{k_{\rm CS}^{(2)}}-\frac{1}{k^{(1)}_{\rm CS}}\right)\Theta(\eta-\eta_*)\right],
\end{equation}
where $\Theta(\eta -\eta_*)$ denotes the Heaviside theta function which vanishes for $\eta<\eta_*$. Note that the transition is abrupt for modes with $k$ smaller than inverse the transition duration, and so eq.~(\ref{eq:two_stage_kcs}) is a good approximation for those modes. Plugging the expression for $\dot{\varphi}$ into eq.~(\ref{eq:ddz_z}) gives
\begin{align}
    \frac{z''_\lambda(\eta, k)}{z_\lambda(\eta, k)} &\simeq \frac{a''}{a} + \frac{a''}{a}\frac{k}{a}\frac{\lambda_{\rm R,L}}{k^{(1)}_{\rm CS}} + \frac{k}{a}\lambda_{\rm R,L} \left(\frac{1}{k^{(2)}_{\rm CS}}- \frac{1}{k_{\rm CS}^{(1)}}\right) \nonumber\\
    &\times \left(\frac{a''}{a}\Theta(\eta-\eta_*)-a'\delta(\eta-\eta_*)-a\delta'(\eta-\eta_*)\right),
\end{align}
and the equation for $X^{(1)}_\lambda$ is now
\begin{equation}
    X^{(1)}_\lambda + \left(k^2-\frac{a''}{a}\right)X^{(1)}_\lambda = J_\lambda(\eta,k),
\end{equation}
with
\begin{widetext}
\begin{align}
    J_\lambda(\eta,k) = \left[\frac{a''}{a}\frac{k}{a}\frac{\lambda_{\rm R,L}}{k^{(1)}_{\rm CS}} + \frac{k}{a}\lambda_{\rm R,L} \left(\frac{1}{k^{(2)}_{\rm CS}}- \frac{1}{k_{\rm CS}^{(1)}}\right)
    \left(\frac{a''}{a}\Theta(\eta-\eta_*)-\delta'(\eta-\eta_*)\right)\right]X^{(0)}(k\eta).
\end{align}    
\end{widetext}
For any $\eta_*$ and any values of $\dot{\varphi}$ before and after the dCS transition, the formal solution for $X_\lambda^{(1)}$ will have the form
\begin{equation}
    X_\lambda^{(1)} = \int d\tilde{\eta}\; G(\eta, \tilde{\eta}) J(\tilde{\eta},k), 
\end{equation}
from which one can find $\beta_\lambda^{(1)}$.

As an illustrative example, for a transition that takes place during radiation domination\footnote{We checked numerically that transitions during matter domination yield negligible corrections to the tensor power spectrum. A motivation for transitions during radiation domination comes from identifying $\phi$ with an axion-like particle (ALP). ALPs often remain frozen at early times due to Hubble friction and start evolving when their mass term becomes dynamically relevant, which can happen during the radiation era depending on the underlying particle physics model (e.g., see the discussion in \cite{Alexander:2024vav}).
}, we have $G(\eta, \tilde{\eta}) = \theta(\eta-\tilde{\eta})k^{-1} \sin k(\eta - \tilde{\eta})$ and $\beta_\lambda^{(1)}$ can be computed straightforwardly for sub-Hubble modes today: 
\begin{align}\label{eq:2stage_dCS_transition}
    \beta_\lambda^{(1)} &\simeq -\lambda_{\rm R,L} \frac{k}{k_{\rm CS}^{(2)}} \frac{\eta_{\rm eq}^2k^2}{a_{\rm eq}}\mathcal{I}(k\eta) +\lambda_{\rm R,L} \frac{k}{a_*} \left(\frac{1}{k^{(2)}_{\rm CS}}-\frac{1}{k^{(1)}_{\rm CS}}\right)\times\nonumber\\
    &\times \left[\frac{i \beta(\eta_{\rm r})}{2k\eta_*}-\alpha(\eta_{\rm r})\left(1-\frac{i}{2k\eta_*}\right)e^{-2ik\eta_*}\right],
\end{align}
where we used the analytical form of $X^{(0)}$ during radiation domination. The first term in the above comes from integration after $\eta_{\rm eq}$ and its relevant limits can be read from \eqref{eq:b1_constdotphi}. More importantly, the second term will dominate the dCS contribution to the energy spectrum that, after inspecting \eqref{eq:total_bsquared}, is proportional to
\begin{align}\label{eq:gamma_2-stage}
    \text{Re}\left[\frac{i}{2k\eta_*}-\frac{\alpha(\eta_{\rm r})}{\beta(\eta_{\rm r})}\left(1-\frac{i}{2k\eta_*}\right)e^{-2ik\eta_*}\right]&\simeq \nonumber\\
    \cos2k(\eta_*- \eta_{\rm r})- \frac{1}{2k\eta_*}\sin2k(\eta_* - \eta_{\rm r})
\end{align}
for super-Hubble modes at $\eta_r$. Hence, the dCS transition induces oscillations in $\rho_{\rm GW
}(k)$ with a frequency fixed by the time when the transition occurs. 

    \begin{figure*}
        \centering
        \includegraphics[width=1\linewidth]{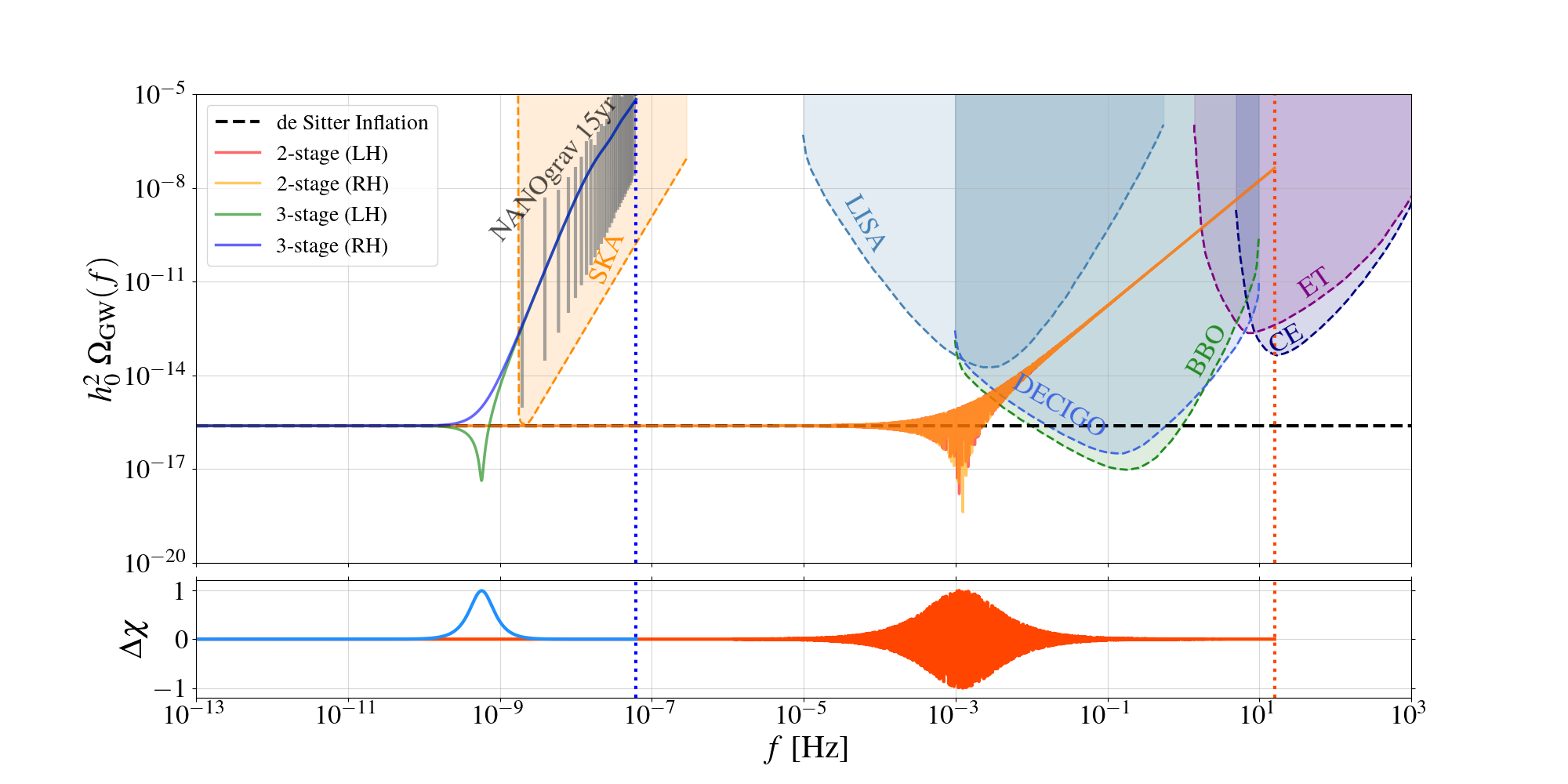}
        \caption{Examples of the power spectrum for two- and  three-stage $\dot{\varphi}$ evolutions, where two and three dCS transitions take place, respectively. The shaded areas are sensitivity curves of the current observations (NANOgrav $15$yr \cite{NANOGrav:2023gor, NANOGrav:2023hvm}) and the sensitivity curves for the future GW experiments: LISA \cite{LISA:2017pwj}, DECIGO \cite{Kawamura:2011zz}, Big Bang Observer (BBO) \cite{Crowder:2005nr}, SKA \cite{Janssen:2014dka}, cosmic explorer (CT) \cite{Reitze:2019iox}, and Einstein telescope (ET) \cite{Punturo:2010zz}. For the two-stage case (in red and orange), we assume $f_{\rm CS}$ transits from $10^{10}~{\rm Hz}$ to $10^5~{\rm Hz}$ at the redshift $z\sim 10^8$. The sudden transition assumption holds only for modes with $k\Delta \eta < 1$. We choose a fiducial cutoff frequency at $\mathcal{O}(10)~{\rm Hz}$. In this example, the left-handed and right-handed modes oscillate with a phase difference, as expected from eq.~(\ref{eq:gamma_2-stage}). The polarization parameter $\Delta \chi$ oscillates accordingly. This particular parameter choice leads to a signal that overlaps with BBO, DECIGO, ET, and CE's sensitivity curves. Different parameter choices also lead to observable signals in LISA. For the three-stage evolution (in blue and green), the CS scalar experiences subsequent static, rolling, and static phases. The static phases, with $\dot\varphi =0$, can model the CS scalar at local minima of its potential. We assume the rolling phase begins at redshift $z\sim 2\times 10^{12}$ and ends at $z\sim 1.8 \times 10^{12}$ with $f_{\rm CS} = 1~\mathrm{Hz}$, and the frequency cutoff is set by $k\Delta\eta <1$. This particular choice offers an alternative explanation of the nanohertz GWs detected by NANOgrav. With a faster rolling speed $\dot\varphi$ and a shorter transition duration, $3$-stage dCS transition can also produce GWs at the higher frequency band.}
        \label{fig:sudden_transition}
    \end{figure*}

Another interesting case is where $\dot{\varphi}$ has a non-vanishing constant value only for a finite period of time, i.e. there are two transitions during the field evolution. This can happen, for example, in a second-order phase transition, when $\varphi$ rolls from an unstable extrema to a local minimum of its potential. To model this transition, we assume the follow expression for $\dot{\varphi}$
\begin{equation}
    \frac{\lambda_{R,L}\dot{\varphi}}{2} = \frac{\lambda_{\rm R,L}}{k_{\rm CS}} \left[\Theta(\eta-\eta_1)- \Theta(\eta-\eta_2)\right],
\end{equation}
where $\eta_2<\eta_{\rm eq}$, i.e., the phase transition happens during radiation domination. In this case, there will be no contribution from the convolution of the Green's function with the source during matter domination (first term in \eqref{eq:2stage_dCS_transition}). Explicitly, we have, for super-Hubble modes today
\begin{equation}
    \beta^{(1)}_\lambda \simeq \lambda_{\rm R,L} \frac{k}{k_{\rm CS}} \left[\frac{i\beta(\eta_r)}{2ka\eta}-\alpha(\eta_{\rm r})\left(1-\frac{i}{2k\eta}\right)\frac{e^{-2ik\eta}}{a(\eta)}\right]_{\eta_2}^{\eta_1}.
\end{equation}
If the total period in which $\varphi$ is evolving is very short, we can write $\eta_1 = \eta_*$ and $\eta_2 = \eta_* + \delta \eta$ where $\delta\eta \ll \eta_*$. In this limit and for super-Hubble modes at $\eta_{\rm r}$, we a contribution to the power spectrum proportional to
\begin{align}
    &\left[\frac{i}{2ka\eta}-\frac{\alpha(\eta_{\rm r})}{\beta(\eta_r)}\left(1-\frac{i}{2k\eta}\right)\frac{e^{-2ik\eta}}{a(\eta)}\right]_{\eta_*+\delta\eta}^{\eta_*} \simeq \nonumber\\
    &2\lambda_{\rm R,L} \frac{k}{k_{\rm CS}}\frac{k}{a(\eta_*)}\delta\eta\bigg(\sin2k(\eta_*-\eta_{\rm r}) - \nonumber\\
    & \left. -\frac{\cos2k(\eta_*-\eta_{\rm r})}{k\eta_*}-\frac{\sin2k(\eta_*-\eta_{\rm r})}{2k^2 \eta_*^2}\right),
\end{align}
which has a different $k$ dependence compared to \eqref{eq:gamma_2-stage}.

\subsection*{Case IV: Constant Equation of State}\label{sec:constantw}
In this section, we allow the Chern-Simons pseudo-scalar $\varphi$ to be dynamical and parametrize its time evolution as an effective fluid. One can use the energy density and pressure of the scalar
\begin{align}
    \rho_\varphi = \frac{\dot \varphi^2}{2} + V, \quad P_\varphi =  \frac{\dot \varphi^2}{2} - V
\end{align}
to obtain $\dot\varphi = \sqrt{\rho_\varphi (1+w_\varphi)}$. The equation of state $w_\varphi \equiv P_\varphi / \rho_\varphi$ is assumed to be constant. The pseudo-scalar energy density $\rho_\varphi \propto \rho_0 a^{-3(w_\varphi+1)}$ redshifts as power law, and the subscript $0$ indicates the pseudo-scalar energy density today. 

With an expression for $\dot\varphi$, we have 
\begin{align}
    z_\lambda (\eta, k) = a(\eta)\sqrt{1- 2\lambda_{L,R} \left(\frac{k}{k_{\rm CS}}\right)a^{-n}}
\end{align}
where $n = (3w_\varphi+5)/2$. For large modes with $(k/k_{\rm CS})a^{-n} \ll 1$, the effective potential becomes
\begin{align}\label{eq:z2z_constantw}
    \frac{z_\lambda''}{z_\lambda} \approx \frac{a''}{a} + n
\lambda_{R,L} \left(\frac{k}{k_{\rm CS}}\right)\left[(1-n)\left(\frac{a'}{a}\right)^2 + \frac{a''}{a}\right]a^{-n}.
\end{align}
For simplicity, we assume that the Chern-Simons modified effect becomes relevant during the radiation-dominated epoch. Starting from the scale-invariant and parity-even power spectrum given by eq.~(\ref{eq:GR_power_spec}), we solve the Bogoliubov coefficients for the matter-dominated epoch in two scenarios: $\emph{(i)}$ for matter-like $\varphi$ evolution $(w_\varphi = 0)$; and $\emph{(ii)}$ for radiation-like $\varphi$ evolution $(w_\varphi = 1/3)$. These equations of state can be obtained from anharmonic oscillations in the minima of renormalizable potentials \cite{Turner:1983he}. We neglect the dark energy case with $w=-1$ for it retains the GR case by setting $\dot\varphi = 0$ and the parity-violating effect is proportional to $\dot\varphi.$ 

The source term for $X_\lambda^{(1)}$ becomes 
\begin{align}\label{eq:source_for_eos}
    J_\lambda(\eta,k) = \Bigg[\lambda_{R,L} \left(\frac{k}{k_{\rm CS}}\right)n\left((1-n)\frac{{a'}^2}{a^2} + \frac{a''}{a}\right)a^{-n}\Bigg]X_\lambda^{(0)}.
\end{align}
Unlike the constant $\dot\varphi$ scenario, the source term in the radiation-dominated epoch does not vanish and is substantially enhanced by $a^{-3}$ ($a^{-5/2}$) for $w_\varphi = \frac{1}{3}$ ($w_\varphi = 0$). Thus, the particle production is sensitive to when the dCS coupling is turned on. 

\subsubsection*{Radiation-Dominated Epoch}
During radiation-dominated epoch, the background solution for $X$ is nothing but free oscillation 
\begin{align}
    X^{(0)}(\eta) = \alpha_{\rm r} e^{-ik\eta} + \beta_{\rm r} e^{ik\eta}.
\end{align}
Using the Green's function method as in the previous sections, but for the source \eqref{eq:source_for_eos}, the leading-order correction to the Bogoliubov coefficients are given by
\begin{subequations}
    \begin{align}
    \alpha_\lambda^{(1)} & = \frac{i}{2} \lambda_{R,L} \left(\frac{k}{k_{\rm CS}}\right)(n-n^2)\left(\frac{k\eta_{\rm eq}}{a_{\rm eq}}\right)^n {\cal I}_{\alpha}^{(\rm RD)}, \\
    \beta_\lambda^{(1)} & = -\frac{i}{2} \lambda_{R,L} \left(\frac{k}{k_{\rm CS}}\right)(n-n^2)\left(\frac{k\eta_{\rm eq}}{a_{\rm eq}}\right)^n {\cal I}_{\beta}^{(\rm RD)},
\end{align}
\end{subequations}
where
\begin{subequations}
    \begin{align}
    {\cal I}_{\alpha}^{(\rm RD)}(x) &= \int_{x_{\rm CS}}^x d\tilde{x}~\left(\alpha_{\rm r} + \beta_{\rm r}e^{2i\tilde{x}}\right) \tilde{x}^{-n-2}, \\
    {\cal I}_{\beta}^{(\rm RD)}(x) &= \int_{x_{\rm CS}}^x d\tilde{x}~\left(\alpha_{\rm r} e^{-2i\tilde{x}} + \beta_{\rm r}\right) \tilde{x}^{-n-2}.
\end{align}
\end{subequations}
\begin{widetext}
For radiation-like ($n=3$) scalar evolution,
    \begin{align}
    \mathcal{I}_{\alpha_{\rm RD}}^{\rm (rad)}(k\eta) &=  \left[\frac{\beta_{\rm r}}{12}\left(\frac{\left(4 i \tilde{x}^3+2 \tilde{x}^2-2 i \tilde{x}-3\right)e^{2 i \tilde{x}}}{\tilde{x}^4}+8 \operatorname{Ei}(2 i\tilde{x})\right) - \frac{\alpha_{\rm r}}{4\tilde{x}^4}\right]_{x_{\rm CS} = k\eta_{\rm CS}}^{x = k\eta},  \\
    \mathcal{I}_{\beta_{\rm RD}}^{\rm (rad)}(k\eta) &=  \left[\frac{\alpha_{\rm r}}{12}\left(\frac{(-4i\tilde{x}^3 + 2\tilde{x}^2 +2i\tilde{x} - 3)e^{-2i\tilde{x}}}{\tilde{x}^4} + 8{\rm Ei}(-2i\tilde{x})\right) - \frac{\beta_{\rm r}}{4\tilde{x}^4}\right]_{x_{\rm CS} = k\eta_{\rm CS}}^{x = k\eta}.
\end{align}
For matter-like ($n=5/2$) scalar evolution 
\begin{align}
    \mathcal{I}_{\alpha_{\rm RD}}^{\rm (mat)}(k\eta) &= \left[\frac{2\beta_{\rm r}}{105}\left(\frac{\left(64 i \tilde{x}^3+16 \tilde{x}^2 -12 i\tilde{x} -15\right)e^{2i\tilde{x}} }{\tilde{x}^{7/2}} - 64\sqrt{2i}\Gamma\left[\frac{1}{2}, -2i\tilde{x}\right]\right) - \frac{2\alpha_{\rm r} }{7\tilde{x}^{7/2}} \right]_{x_{\rm CS} = k\eta_{\rm CS}}^{x = k\eta} \\
    \mathcal{I}_{\beta_{\rm RD}}^{\rm (mat)}(k\eta) &= \left[\frac{2\alpha_{\rm r}}{105}\left(\frac{\left(-64 i \tilde{x}^3+16 \tilde{x}^2 +12 i\tilde{x} -15\right)e^{-2i\tilde{x}} }{\tilde{x}^{7/2}} - 64\sqrt{2i}\Gamma\left[\frac{1}{2}, 2i\tilde{x}\right]\right) - \frac{2\beta_{\rm r} }{7\tilde{x}^{7/2}} \right]_{x_{\rm CS} = k\eta_{\rm CS}}^{x = k\eta} 
\end{align}
\end{widetext}
For super-Hubble modes today and for $k\eta_{\rm CS} < k\eta_{\rm eq} < 1$ the Bogoliubov coefficients become
\begin{align}
    (w_{\varphi} = \frac{1}{3})&~\begin{cases}
        \alpha_\lambda^{(1)} &\to -\frac{3i\lambda_{\rm R,L}}{4}\left(\frac{k}{k_{\rm CS}}\right)\left(\frac{k\eta_{\rm eq}}{a_{\rm eq}}\right)^3\frac{(\alpha_{\rm r} + \beta_{\rm r})}{(k\eta_{\rm CS})^4} \\
        \beta_\lambda^{(1)} &\to \frac{3i\lambda_{\rm R,L}}{4}\left(\frac{k}{k_{\rm CS}}\right)\left(\frac{k\eta_{\rm eq}}{a_{\rm eq}}\right)^3\frac{(\alpha_{\rm r} + \beta_{\rm r})}{(k\eta_{\rm CS})^4}
    \end{cases}
\end{align}
\begin{align}
    (w_{\varphi}=0)&~\begin{cases}
        \alpha_\lambda^{(1)} &\to -\frac{15i\lambda_{\rm R,L}}{28}\left(\frac{k}{k_{\rm CS}}\right)\left(\frac{k\eta_{\rm eq}}{a_{\rm eq}}\right)^{\frac{5}{2}}\frac{(\alpha_{\rm r} + \beta_{\rm r})}{(k\eta_{\rm CS})^{7/2}} \\
        \beta_\lambda^{(1)} &\to ~\frac{15i\lambda_{\rm R,L}}{28}\left(\frac{k}{k_{\rm CS}}\right)\left(\frac{k\eta_{\rm eq}}{a_{\rm eq}}\right)^{\frac{5}{2}}\frac{(\alpha_{\rm r} + \beta_{\rm r})}{(k\eta_{\rm CS})^{7/2}}
    \end{cases}
\end{align}
while for the $k\eta_{\rm eq} > k\eta_{\rm CS} > 1$ limit we have
\begin{align}
    (w_{\varphi}=\frac{1}{3})&~\begin{cases}
        \alpha_\lambda^{(1)} &\to -\frac{3i\lambda_{\rm R,L}}{4}\left(\frac{k}{k_{\rm CS}}\right)\left(\frac{k\eta_{\rm eq}}{a_{\rm eq}}\right)^3\frac{\alpha_{\rm r}}{(k\eta_{\rm CS})^4} \\
        \\
        \beta_\lambda^{(1)} &\to \frac{3i\lambda_{\rm R,L}}{4}\left(\frac{k}{k_{\rm CS}}\right)\left(\frac{k\eta_{\rm eq}}{a_{\rm eq}}\right)^3\frac{\beta_{\rm r}}{(k\eta_{\rm CS})^4}
    \end{cases}
\end{align}
\begin{align}
    (w_{\varphi}=0)&~\begin{cases}
        \alpha_\lambda^{(1)} &\to -\frac{15i\lambda_{\rm R,L}}{28}\left(\frac{k}{k_{\rm CS}}\right)\left(\frac{k\eta_{\rm eq}}{a_{\rm eq}}\right)^{\frac{5}{2}}\frac{\alpha_{\rm r} }{(k\eta_{\rm CS})^{7/2}} \\
        \\
        \beta_\lambda^{(1)} &\to ~\frac{15i\lambda_{\rm R,L}}{28}\left(\frac{k}{k_{\rm CS}}\right)\left(\frac{k\eta_{\rm eq}}{a_{\rm eq}}\right)^{\frac{5}{2}}\frac{\beta_{\rm r}}{(k\eta_{\rm CS})^{7/2}}.
    \end{cases}
\end{align}

\subsubsection*{Matter-Dominated Epoch}

During the matter-dominated epoch, the background $X^{(0)}$ is given by 
\begin{align}
     X^{(0)}(\tau) = b_1 \left(1 - \frac{i}{k\tau}\right)e^{-ik\tau} + b_2\left(1 + \frac{i}{k\tau}\right)e^{ik\tau},
\end{align}
where we define $\tau \equiv \eta + \eta_{\rm eq}$ and the coefficients are
\begin{align}
    b_1 &=  \alpha_\lambda^{(\rm eq)}\left(1+ \frac{i}{k\tau_{\rm eq}}- \frac{1}{2k^2 \tau^2_{\rm eq}}\right)e^{ik\eta_{\rm eq}}-\beta_\lambda^{(\rm eq)}\frac{e^{3ik\eta_{\rm eq}}}{2k^2\tau^2_{\rm eq}}, \\
    b_2 &= -\alpha_\lambda^{(\rm eq)} \frac{e^{-3ik\eta_{\rm eq}}}{2k^2 \tau^2_{\rm eq}} + \beta_\lambda^{(\rm eq)}\left(1 - \frac{i}{k\tau_{\rm eq}}- \frac{1}{2k^2 \tau^2_{\rm eq}}\right)e^{-ik\eta_{\rm eq}}.
\end{align}\\
Since the Bogoliubov coefficients are amplified during RD, we define $\alpha_\lambda^{(\rm eq)} \equiv \alpha_{\rm r} + \alpha_{\lambda}^{(1)}(\eta_{\rm eq})$ and similar to $\beta_\lambda^{(\rm eq)}$. Note that the $O(\alpha)$ evolution during radiation domination was left in $X^{(0)}$, so that the boundary conditions for the Green function is unchanged.
The source term in the matter-dominated epoch is 
\begin{align}
    J_\lambda(\eta,k) = \lambda_{R,L} \left(\frac{k}{k_{\rm CS}}\right)n(6-4n)\left(\frac{4\eta_{\rm eq}^2}{a_{\rm eq}}\right)^n \tau^{-2n-2} .
\end{align}
Since we are interested in the modes within the Hubble radius today, the leading order correction to $\beta$ is computed in the sub-Hubble limit $k \eta_0 \gg 1$
\begin{align}
    \beta_\lambda^{(1)}&=-\frac{1}{2}\lambda_{R,L} \left(\frac{k}{k_{\rm CS}}\right)n(6-4n)\left(\frac{4k^2\eta_{\rm eq}^2}{a_{\rm eq}}\right)^n \mathcal{I}_\beta^{(\rm MD)}.\\
    \mathcal{I}_\beta^{(\rm MD)}&= \int_{k\tau_{\rm eq}}^{k\tau} d\tilde{x} ~\Bigg[b_1 \left(1 - \frac{i}{\tilde{x}}\right)\left(\frac{1+i\tilde{x}}{\tilde{x}^{(2n+3)}}\right)e^{-2i\tilde{x}} \nonumber \\
    & \hspace{1.5cm} + b_2\left(1 + \frac{i}{\tilde{x}}\right)\left(\frac{1+i\tilde{x}}{\tilde{x}^{(2n+3)}}\right) \Bigg].
\end{align}
Notice that the constant $\dot\varphi$ case in can be reproduced by setting $n =1$.

The solution for $\beta^{(1)}_\lambda$ during matter domination the matter-like ($n = 5/2$) and radiation-like ($n = 3$) pseudo-scalar evolution can be read from the integrals
\begin{widetext}
    \begin{align}
    \mathcal{I}_{\beta_{\rm MD}}^{\rm (mat)}(k\eta) &= \Bigg[\frac{ib_1}{360}\Bigg(\frac{\left(45 + 90i\tilde{x} - 30\tilde{x}^2 + 12i\tilde{x}^3 + 6\tilde{x}^4-4i\tilde{x}^5-4\tilde{x}^6+8i\tilde{x}^7\right)}{\tilde{x}^8}e^{-2i\tilde{x}} - 16{\rm Ei}[-2i\tilde{x}]\Bigg) \nonumber\\
    & \hspace{10cm}- ib_2\left(\frac{1}{8\tilde{x}^8}+\frac{1}{6\tilde{x}^6}\right)\Bigg]_{x_{\rm eq} = k\tau_{\rm eq}}^{x = k\tau},  \\
    \mathcal{I}_{\beta_{\rm MD}}^{\rm (rad)}(k\eta) &=\Bigg[\frac{ib_1}{567}\Bigg(\frac{\left(63 + 126i\tilde{x} -45\tilde{x}^2 
    + 15i\tilde{x}^3 + 6\tilde{x}^4 - 3i\tilde{x}^5 -2\tilde{x}^6 + 2i\tilde{x}^7 +4\tilde{x}^8 \right)}{\tilde{x}^{9}}e^{-2i\tilde{x}} \nonumber\\
    & \hspace{7cm} + 8i{\rm Ei}[-2i\tilde{x}]\Big) - ib_2\left(\frac{1}{9\tilde{x}^{9}}+\frac{1}{7\tilde{x}^7}\right)\Bigg]_{x_{\rm eq} = k\tau_{\rm eq}}^{x = k\tau}.
    \end{align}
\end{widetext}
In the sub-Hubble limit today ($k\eta \gg 1$), and for $k\eta_{\rm eq} \gg 1$ we have
\begin{align}
    \beta_\lambda^{(1)} \to 
    \begin{cases}
        \frac{9i\lambda_{\rm R,L}}{7}\left(\frac{k}{k_{\rm CS}}\right)\left(\frac{4k^2\eta_{\rm eq}^2}{a_{\rm eq}}\right)^3\frac{b_2}{(2k\eta_{\rm eq})^7}, \quad (w_{\varphi}=\frac{1}{3}) \\
        \\
        \frac{5i\lambda_{\rm R,L}}{6}\left(\frac{k}{k_{\rm CS}}\right)\left(\frac{4k^2\eta_{\rm eq}^2}{a_{\rm eq}}\right)^{\frac{5}{2}}\frac{b_2}{(2k\eta_{\rm eq})^6}, \quad (w_{\varphi}=0)
    \end{cases},
\end{align}
or taking the limits of the coefficients,
\begin{align}
    \beta_\lambda^{(1)} \to 
    \begin{cases}
        \frac{9i\lambda_{\rm R,L}}{7}\left(\frac{k}{k_{\rm CS}}\right)\left(\frac{4k^2\eta_{\rm eq}^2}{a_{\rm eq}}\right)^3\frac{\beta_\lambda^{(\rm eq)}e^{-ik\eta_{\rm eq}} }{(2k\eta_{\rm eq})^7}, \quad (w_{\varphi}=\frac{1}{3})\\
        \\
        \frac{5i\lambda_{\rm R,L}}{6}\left(\frac{k}{k_{\rm CS}}\right)\left(\frac{4k^2\eta_{\rm eq}^2}{a_{\rm eq}}\right)^{\frac{5}{2}}\frac{\beta_\lambda^{(\rm eq)}e^{-ik\eta_{\rm eq}}}{(2k\eta_{\rm eq})^6}, \quad (w_{\varphi}=0)
    \end{cases}
\end{align}
from which one can see there is an oscillatory behaviour. For $k\eta_{\rm eq}\ll 1$, we have
\begin{align}
    \beta_\lambda^{(1)} \to 
    \begin{cases}
        i\lambda_{\rm R,L}\left(\frac{k}{k_{\rm CS}}\right)\left(\frac{4k^2\eta_{\rm eq}^2}{a_{\rm eq}}\right)^3\frac{b_2-b_1}{(2k\eta_{\rm eq})^9}, \quad (w_{\varphi} =\frac{1}{3}) \\
        \\
        \frac{5i\lambda_{\rm R,L}}{8}\left(\frac{k}{k_{\rm CS}}\right)\left(\frac{4k^2\eta_{\rm eq}^2}{a_{\rm eq}}\right)^{\frac{5}{2}}\frac{b_2-b_1}{(2k\eta_{\rm eq})^{8}}, \quad (w_{\varphi}=0)
    \end{cases}
\end{align}
or
\begin{align}
    \beta_\lambda^{(1)} \to 
    \begin{cases}
        \frac{2\lambda_{\rm R,L}}{3}\left(\frac{k}{k_{\rm CS}}\right)\left(\frac{4k^2\eta_{\rm eq}^2}{a_{\rm eq}}\right)^3\frac{\alpha_\lambda^{(\rm eq)} +\beta_\lambda^{(\rm eq)}}{(2k\eta_{\rm eq})^8}, \quad (w_{\varphi} =\frac{1}{3}) \\
        \\
        \frac{5\lambda_{\rm R,L}}{12}\left(\frac{k}{k_{\rm CS}}\right)\left(\frac{4k^2\eta_{\rm eq}^2}{a_{\rm eq}}\right)^{\frac{5}{2}}\frac{\alpha_\lambda^{(\rm eq)} +\beta_\lambda^{(\rm eq)}}{(2k\eta_{\rm eq})^{7}}, \quad (w_{\varphi}=0)
    \end{cases}
\end{align}

\begin{figure*}[t]
        \centering
        \includegraphics[width=1\linewidth]{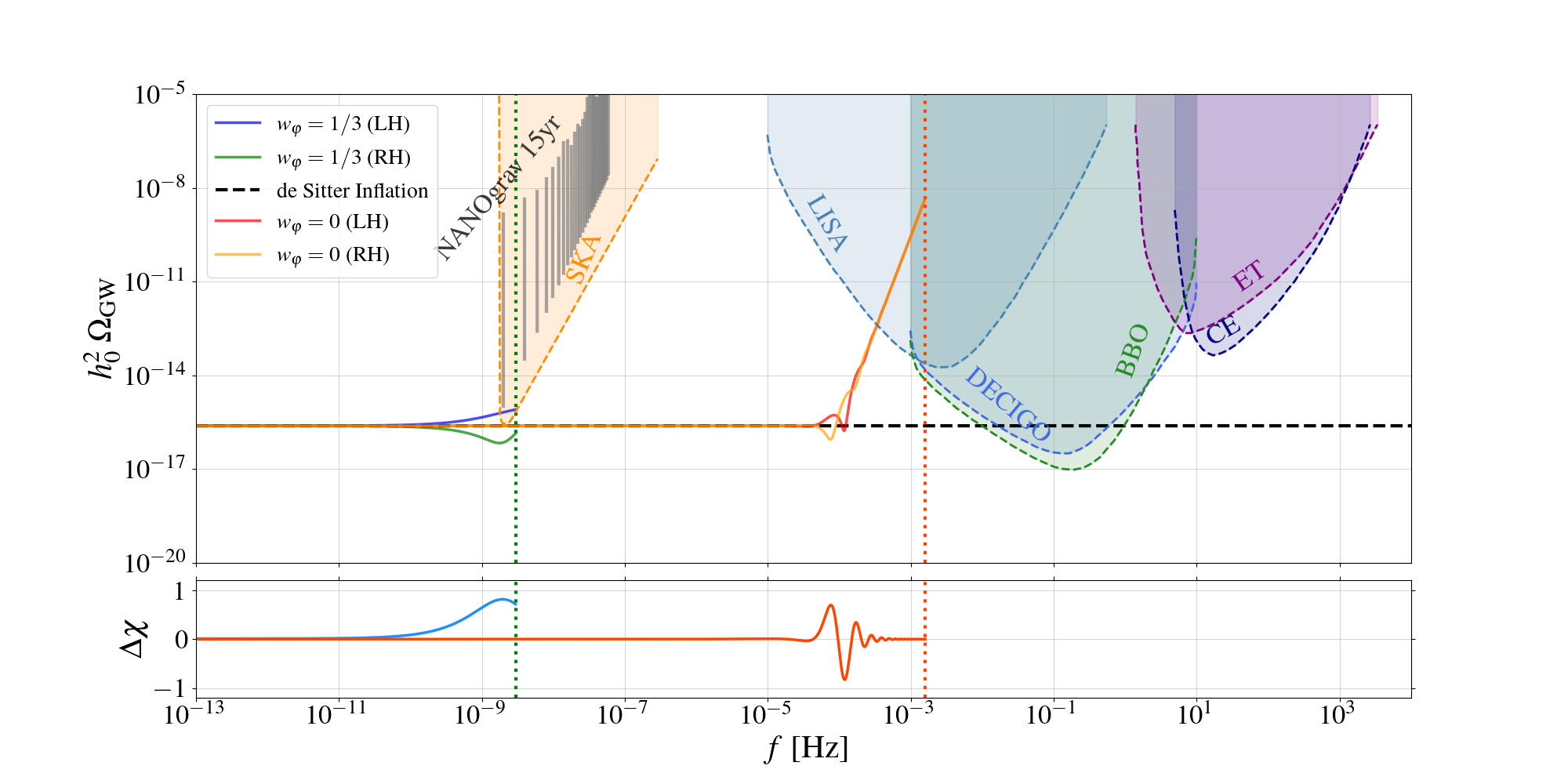}
        \caption{Examples of power spectra of the radiation-like and matter-like dCS induced gravitational waves. The shaded areas are the same as figure~\ref{fig:sudden_transition}. For $w_{\varphi} = 0$ (in red and orange), we assume the Chern-Simons psuedo-scalar becomes dynamical at the redshift $z\sim 10^{16}$. The energy density $\rho_\varphi$ is chosen such that $f_{\rm CS} = 10^{38}~{\rm Hz}$. Since this work focus only on the perturbative region, \textit{i.e.}, $(f/f_{\rm CS}) \ll 1$, the cutoff frequency is set based on eq.~(\ref{eq:cutoff-freq}). For the $w_{\varphi} = 1/3$ case, we assume the CS scalar becomes dynamical at $z\sim 1.5 \times 10^{12}$ with a constant $f_{\rm CS} = 10^{28}~{\rm Hz}$. A fully polarized GW spectrum overlaps with the sensitivity curve of SKA. The vacuum amplification would also imprint on the non-perturbative region where $f> f_{\rm CS}$, and we leave this for future studies. }
        \label{fig:constant_w}
    \end{figure*}

Due to the $a^{-n}$ dependence of the CS factor in $z_\lambda$, the vacuum amplification effect is rapidly washed out by redshifting. During matter-domination, the extra particle production comparing to GR is suppressed by $k\eta_{\rm eq}$ and $a_{\rm eq}$. Hence, the power spectrum for ultra-low frequency band is identical to the pure GR case (see figure~\ref{fig:constant_w}).

\section{Discussion and Conclusion}\label{sec:conclusion}

Research on properties of the stochastic background of gravitational waves has been a tool for studying general relativity and gravity theories beyond GR. In this paper, we focused on graviton production via vacuum amplification in dynamical Chern-Simons gravity theory. We have laid out the formalism to find graviton production in dCS by using a timely continuous Bogoliubov transformation to track the evolution of an initially vacuum state in cosmology. We studied the vacuum amplification of GWs in dCS for four different scenarios and investigated their possible imprint on the SGWB energy spectrum that can be measured in current and future observations. 

After a review of vacuum amplification in GR, we have found the gravitational energy under different transitions between cosmic epochs. We then provided the formalism for vacuum amplification in dCS gravity, which results in a polarized energy spectrum of SGWB. We defined a physical scale $k_{\rm CS}$ in terms of the CS pseudoscalar field $\varphi$ (see eq. \eqref{kCS}) and showed that the dCS contribution of the energy spectrum is proportional to $k/k_{\rm CS}$ (see eq. \eqref{eq:dCS_rho_k}).

We applied the framework developed in section \ref{sec:vacuum_amp_CS} to four scenarios where we explored the vacuum amplification in dCS gravity theory: the Minkowski background limit, constant $\dot{\varphi}$ (non-dynamical CS gravity), constant equation of state parameterization of $\varphi$ dynamics, and dCS transitions. For reasons we discuss in the following, the latter two cases provided power spectra that can be probed by observations.  

In Minkowski limit case, we ignore the cosmic expansion but still consider the SGWB to be cosmological. Under the assumption that $\varphi$ has a mass potential, the metric mode function undergoes parametric resonance due to $\varphi$ oscillations. However, in the perturbative regime of the frequency space explored, within the instability bands of the resonance, parity-even terms dominate over parity-odd ones in the expression for the GW power spectrum. This is due to the scalar energy density being much smaller than the critical density, and departing from this assumption might give a parity-odd spectrum within the perturbative regime, as is the case of having astrophysical sources of SGWB, for which the pseudo-scalar field energy might be larger enough to have observational parity violating signal. We have also considered non-dynamical CS gravity, for which $\dot{\varphi}$ is constant. In this case, the modification of the GW energy spectrum in GR is negligibly small. Here, the size of the CS contribution to the spectrum is bounded by astrophysical constraints on $\dot{\varphi}$. 

We also considered the scenario where $\dot{\varphi}$ changes between two constant asymptotic behaviors or when it has a constant, non-vanishing value only for a finite period of time (we call these evolutions dCS transitions). The latter can approximate the time evolution of $\varphi$ during a phase transition, when it rolls towards a possible new local minima of its potential. We considered transitions that occur during matter or radiation domination. The power spectrum for both of these examples is presented in Fig. \ref{fig:sudden_transition}, and again they intersect with observationally relevant regions. The spectra are both strongly polarized, with the rolling example having phase-different oscillations for right- and left-handed polarizations.

Another scenario we have studied was the assumption of an effective fluid approach for the time evolution of the dynamical pseudo-scalar field $\varphi$. We parametrized it by constant equations of state, for which we let it to be radiation- or dust-like. We computed the power spectrum for vacuum amplification when the field becomes dynamical during radiation- and matter-dominated epochs, both for $\varphi$'s energy density redshifting as radiation or dust. We have shown that the effect of vacuum amplification is quickly washed out by redshifting. However, for values of $k_{\rm CS}$ small enough, the power spectrum today is larger than the inflationary one within the perturbative frequency range explored. In Fig. \ref{fig:constant_w} we present the power spectrum of GW for this scenario, and the result intersects the observational region (sensitivity curve) for some GW observational searches. For some frequencies, there is a significant difference between the left- and right-handed spectra, depicting the parity violation nature of CS gravity. We have only considered a perturbative analysis on the CS contribution, and we leave further studies on the non-perturbative frequency region for the future.

The formalism described in section \ref{sec:vacuum_amp_CS} is general enough to accommodate different initial spectra and cosmological evolution. The choice of the initial spectrum as the flat one from the dS inflation was a simplified assumption that can be directly extended to accommodate more complex spectra. Moreover, for generality, we have not specified the microphysical origin of the chosen time dependence of $\varphi$. In fact, our framework can be applied for any evolution, and the energy-power spectrum can be used as a probe for the pseudoscalar potential. From this perspective, SGWB observations can be used to constrain the evolution of $\varphi$ within dCS gravity. We hope that the framework discussed in this paper inspires more studies on numerical investigations on the non-perturbative frequency region and, more generally, on parity-violating effects for SGWBs in other modified gravity theories. 

\section*{Acknowledgments}

We thank Robert Brandenberger, Tucker Manton, and Nicol\'as Yunes and for discussions and comments on an early version of the manuscript. This work was supported by the Simons Foundation through Award No. 896696. 

\bibliographystyle{apsrev4-2}
\bibliography{ref}

\end{document}